\documentclass[aps,pra,floatfix,showpacs,preprintnumbers,amsmath,amsymb,a4paper,twocolumn,amsfonts]{revtex4-1}
\usepackage{adjustbox}

\usepackage[utf8]{inputenc}
\usepackage[T1]{fontenc}

\usepackage{amssymb}
\usepackage{amsmath}
\usepackage{graphicx}
\usepackage{color}
\usepackage[colorlinks=true,linkcolor=blue,citecolor=red,urlcolor=blue]{hyperref}

\usepackage{enumitem,amssymb}
\newlist{todolist}{itemize}{2}
\setlist[todolist]{label=$\square$}
\usepackage{pifont}

\newcommand{\be}{\begin{equation}}
\newcommand{\ee}{\end{equation}}
\newcommand{\dd}[1]{\mathrm{d}{#1}}

\newcommand{\ddn}[2]{\mathrm{d^{#1}}{#2}}

\def\i{\text{i}}

\begin{document}
\title{Rydberg Composites} 
%\title{\blue{Spatial Confinement And Geometric Effects In The Electronic Structure
%Of A Rydberg Atom Embedded In A Neutral Medium} \red{needs revision / new title?}}
\author{Andrew~L. Hunter}
\author{Matthew~T. Eiles}
\author{Alex~Eisfeld}
\author{Jan~M. Rost}
\affiliation{ Max Planck Institute for the Physics of Complex Systems, 38 N\"{o}ethnitzer Str., Dresden, Germany}
\date{\today}

\begin{abstract} We introduce the Rydberg Composite, a new class of Rydberg matter where a single Rydberg atom is interfaced with a dense environment of neutral ground state atoms.  The properties of the Composite depend on both the Rydberg excitation, which provides the gross energetic and spatial scales, and on the distribution of ground state atoms within the volume of the Rydberg wave function, which sculpt the electronic states. The latter range from the “trilobites,'' for small numbers of scatterers, to delocalized and chaotic eigenstates for disordered scatterer arrays, culminating in the dense scatterer limit in symmetry-dominated wave functions which promise good control in future experiments.  We characterize these scenarios with
%a variety of
different theoretical methods, enabling us to obtain scaling behavior for the regular spectrum and measures of chaos and delocalization in the disordered regime.
%, and accurate characterizations
Thus, we obtain a systematic description of the Composite states.  The $2$D monolayer Composite possesses the richest spectrum with an intricate band structure in the limit of homogeneous scatterers. 

\end{abstract}
\maketitle

\section{Introduction}
Ultra long-range molecules composed of a Rydberg atom and a ground state atom, colloquially known as trilobites, were proposed in  2000 \cite{Greene2000}. Soon thereafter theoretical explorations regarding the possibility of polyatomic molecules involving several ground state atoms followed \cite{Rost2006,Rost2009}. The experimental verification of Rydberg molecules in 2009 \cite{Bendkowsky} also confirmed accidentally the existence of trimers \cite{quantumreflection}. Since then, interest in Rydberg excitations beyond isolated atoms has rapidly 
branched out into quite diverse scenarios. These include the replacement of the ground state atom in the original trilobite dimer by larger and more complex systems, e.g., one or more polar molecules \cite{Seth,Sadeghpour, GFerezNew,PhysRevA.85.052511}, the (re-)discovery of Rydberg excitations in solid state systems \cite{CupOxide}, and a large variety of excitonic Rydberg dynamics in the gas phase \cite{schauss2012observation,PhysRevA.76.013413,PhysRevLett.104.223002}, just to name a few.  For the increasingly dense gases now achievable in experiments, one can elegantly describe this system as a Rydberg excitation dressed by ground state atoms from the gas. In fact, recent experiments exhibit spectral features corresponding
to polyatomic molecules containing up to five ground state atoms \cite{WhalenPoly,Whalen2,NatCommPfau}, and mean-field shifts in the spectrum reveal this polaronic behavior involving the coupling of many hundreds of atoms to the Rydberg electron \cite{Demler}.  One may wonder how many ground state ``scatterer'' atoms  within the volume occupied by the Rydberg wave function can a trilobite molecule tolerate. A recent study found that under certain conditions the formation of trilobites actually thrives in a dense gas, which is counter-intuitive at first glance \cite{perttu}.

What is lacking is a systematic approach which connects the trilobite regime with a few scatterers to the regime of very dense scatterers, although the phenomena just described
suggest that Rydberg excitations immersed in dense and structured media might have very interesting properties. The present investigation opens a new venue for Rydberg Composite systems along this way, which involve many hundreds of atoms in a structured environment coupled to a single Rydberg atom. These Composites can be formed, for example, by exciting an atom in a 1, 2, or 3-dimensional optical lattice to a Rydberg state which envelops many atoms on surrounding sites. We present a systematic and detailed investigation of this Rydberg Composite  and provide its properties as a function of principal quantum number $\nu$, lattice constant $d$, and fill factor $F$ of lattice sites. 

With the Rydberg composite we change the perspective from the molecular one -- using chemical approaches to characterize polyatomic trilobites via Born-Oppenheimer potential surfaces, rovibrational couplings, etc \cite{EilesJPB,Rost2006} -- to a condensed matter one, emphasizing generic scaling principles, gross structure, and properties associated with the high density of states obtained here. This allows us to approach systematically dense atomic environments. Indeed, 
we will see that towards the limit of homogeneous filling a band-like structure of the spectrum emerges. Moreover, the unique property of a Rydberg electron bound to an isolated atom with a singular point of infinite density of states (DoS) at the ionization threshold $\lim_{\nu\to\infty}E_{\nu}\equiv -1/(2\nu^{2})=0$ and full degeneracy makes such a  Rydberg Composite an interesting object to study, as the distribution of scatterers can break the degeneracy in a controlled, yet flexible, way. We will identify non-trivial scaling properties as a function of $\nu$. They allow us to
connect the situation at finite $\nu$  with threshold $\nu\to\infty$. Finally, the Composite's key properties are derived analytically in the homogeneous limit, while random matrix theory is used for the irregular part of the spectrum. 

We will also explain how  a planar environment  breaks the symmetry of the  Rydberg Composite and leads to much richer spectral structures as compared to
a wire-like (one-dimensional) or crystal-like (three-dimensional) atomic environment. Hence,  we put emphasis on a planar sheet of atoms arranged in a lattice containing a Rydberg excitation as an exemplary Rydberg Composite whose experimental realization is facilitated by the routine creation of two-dimensional optical lattices \cite{bloch2005ultracold}.

This paper is structured as follows.  Sec. \ref{sec:theory} provides the theoretical background. In \ref{sec:generic} we introduce a generic Hamiltonian which represents a broad class of systems consisting of an excited object coupled to localized scatterers. Sec. \ref{sec:specific} defines our specific realization of this Hamiltonian: the Rydberg Composite in $D$ dimensions. Sec. \ref{sec:latticeproperties} details the scaling properties of this Composite system in one, two, and three dimensions. In Sec. \ref{sec:phenomenology} we introduce the phenomenology of the Composite in the three different lattice geometries, investigating both the DoS (Sec. \ref{sec:dos}) and exemplary wave functions (Sec. \ref{sec:wavefunctions}).  In section \ref{sec:homogeneous} we focus on the homogeneous density regime where the system can be studied analytically to obtain a clear intuitive picture of the system, its band structure, and the resulting scaling laws. Section \ref{sec:evolution} investigates the inhomogeneous regime, using statistical measures derived from random matrix theory to reveal that it exhibits quantum chaos. Sec. \ref{sec:experiment} discusses potential experimental realizations, and Sec. \ref{sec:conclusion} concludes with further perspectives and implications. Throughout we adopt atomic units.

\section{Theoretical Description}
\label{sec:theory}
\subsection{Generic Hamiltonian}
\begin{figure*}[t]
    \centering
    \includegraphics[width=\textwidth]{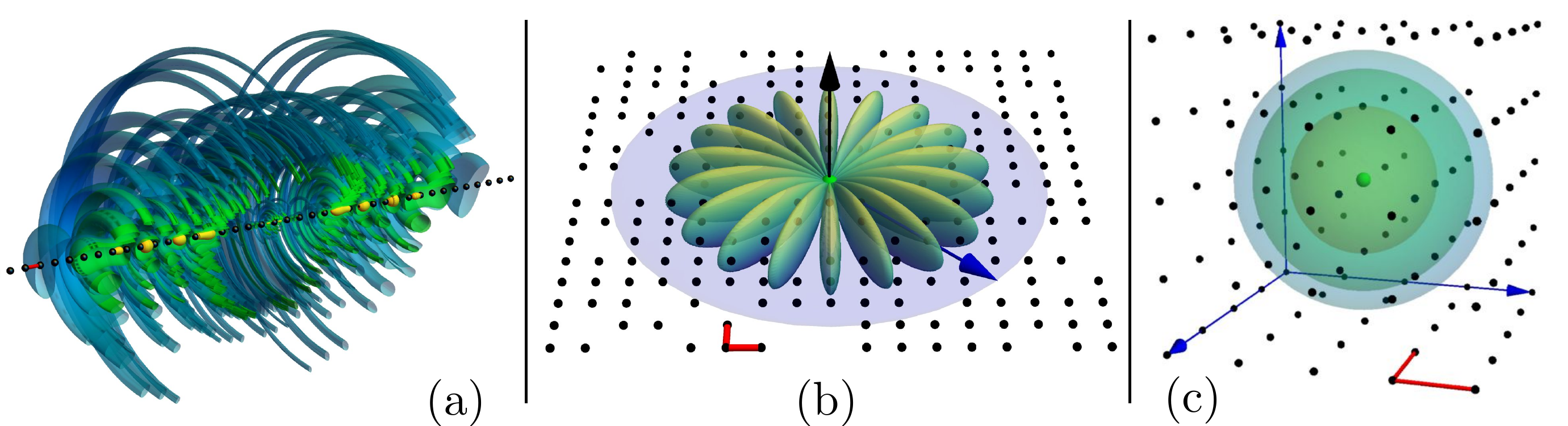}
    \caption{Schematics of the three scenarios we consider: (a) a linear chain ($1$D), (b) a monolayer ($2$D), and (c) a cubic lattice ($3$D). In each panel the black spheres represent scatterers sitting on lattice sites and the red lines give the lattice spacing, $d$. The missing scatterers in (b) represent a situation with non-unity filling.  The volume of scatterers situated within the Rydberg wave function is represented by the blue circle in (b) and by the sphere in (c). The exemplary densities shown give different representations of the Composite's electronic wave function in these three scenarios. In (a) the strongly perturbed wave function is shown with three surfaces of constant density, revealing the exotic nature of these wave functions. The full $3$D contour is cutaway in front to reveal the interior structure. Fig.~\ref{fig:1dwfs} provides details of the  plot parameters. Panel (b) shows a contour of the angular dependence of a typical circular state, which plays a crucial role in the $2$D-Composite's properties (see Sec. \ref{sec:homogeneous}). Panel (c) shows a cartoon of the Rydberg atom, illustrating that the spatially-varying probability cloud spans many lattice sites of the $3$D lattice.  }
    \label{fig:introschematic}
\end{figure*}
\label{sec:generic}
We begin with a generic description of our system, which is composed of an electron with position $\vec r$ and momentum $\vec p$ in the presence of a central potential $V(r)$ and a collection of point-like scattering objects with positions following a distribution of particles $\rho(\vec x)$. This scatterer arrangement can correspond to either a structured geometry or a disordered environment, i.e. that found naturally in an optical lattice or in an ultracold gas, respectively.  Although the electron wave function is fully three-dimensional ($3$D), the dimensionality of the scatterer geometry can be lower, for example as in a one-dimensional ($1$D) chain or a ($2$D) random gas. The scatterers interact with the electron via the potential $U(\vec x,\vec r)$, which destroys the spherical symmetry of the central potential $V(r)$ and, in general, makes the system classically chaotic.
%$U(\vec x,\vec r)$ is perturbative relative to $V(r)$. 
We assume a frozen-gas scenario, consistent with the ultracold temperatures of such a system, and neglect the motion of these scatterers. The electronic Hamiltonian is therefore
\be
\label{eq:genericH}
H=\frac{\vec p^2}{2m} + V( r) + \int \rho(\vec x)U(\vec x,\vec r)\ddn{3}{x}.
\ee

This generic Hamiltonian has been studied in several contexts over the past decades, with examples ranging from two-dimensional ($2$D) quantum dots \cite{prbperttu,rahkonnen2019}, quantum billiards \cite{pre74}, Coulomb systems \cite{sitenko,baltenkov},  perturbed harmonic oscillators \cite{Demiralp1}, and Bose-Einstein condensates in a dimple potential \cite{Demiralp2}, to name just a few. The electronic wave function for vanishing $U$ separates in spherical coordinates: $\Psi(\vec r) = \langle \vec r|\nu lm\rangle = \frac{u_{\nu l}(r)}{r}Y_{lm}(\theta,\phi)$, where $Y_{lm}(\theta,\phi)$ is a spherical harmonic. Therefore, mutual eigenstates of the angular momentum operator and the Hamiltonian satisfy $\vec L^2|\nu lm\rangle = l(l+1)|\nu lm\rangle$, $L_z|\nu lm\rangle = m|\nu lm\rangle$, and $H|\nu lm\rangle = E_{\nu l}|\nu lm\rangle$. Any central potential possesses azimuthal symmetry and hence has $2l-1$ degenerate $|\nu l\rangle$ states. In the next section we consider the Coloumb potential, $V(r)=-1/r$, which has 
%A crucial component of our study is that we choose a $V(r)$ possessing
an additional symmetry: it conserves the Runge-Lenz vector $\vec{A}=\vec{p}\times \vec{L} -\hat r$,
leading to a particularly large degenerate Hilbert space in each manifold $\nu$.
Scatterers will lift this degeneracy. 
Special scatterer geometries, however, may be able to restore
%and through various choices of the scatterer geometry we can dismantle and perhaps even reconstruct 
this degeneracy in the Rydberg Composite.

\subsection{Implementation for the Rydberg Composite}

\label{sec:specific}
The Rydberg atom is a major workhorse of modern atomic physics; here, when embedded in an ultracold medium  of neutral atoms, it provides an ideal physical realization of the Hamiltonian (Eq.~\ref{eq:genericH}). For an alkali atom, this means that $V(r)$ is a Coulomb potential for $r$ larger than a few atomic units. The deviation at small $r$, typically set by an empirical model potential, includes the interactions with the other atomic electrons. This leads to energies $E_{\nu l}$ that are non-degenerate for different $l$ values and noninteger values of $\nu$, the principle quantum number. However, as $l$ increases, the wave function's overlap with this short range region decays rapidly and $E_{\nu l}\to-\frac{1}{2\nu^2}$, where $\nu$ is an integer. Typically only the states with the three or four lowest $l$ values deviate appreciably from the hydrogenic Rydberg spectrum.  The overwhelming majority of states behave as in hydrogen, and therefore for simplicity we consider only a hydrogenic spectrum here with $\nu$ an integer.   For the interaction between the surrounding ultracold atoms -- the localized scatterers -- and the electron we use the Fermi pseudopotential \cite{Greene2000,Fermi}, 

\be
\label{eq:fermi}
U(\vec x,\vec r) = 2\pi a_s(k_{\vec x})\delta^3(\vec x - \vec r) = 2\pi a_s(k_{\vec x})|\vec x\rangle\langle \vec x|,
\ee 
which is straightforward to implement and manipulate. The strength of each scatterer's contribution is given by the energy dependent s-wave electron-atom scattering length $a_{s}[k_{\vec x}]$ \cite{bahrim2000low}. 
This simple contact potential is a reasonable approximation since 
 a neutral atom in its ground state is a highly localized and isotropic perturbation when compared to the Coulomb potential and Rydberg wavelength. It therefore imparts only an $s$-wave phase shift onto the Rydberg wave function via elastic scattering, as characterized by the scattering length. The energy dependence of this process is set by the semiclassical electron momentum, $k_{x}^2 = -\frac{1}{\nu^2} + \frac{2}{|\vec x|}$.

For the Rydberg Composite, Eq.~\ref{eq:genericH} reads
\begin{subequations}
\be
\label{eq:ham}
    {H}=-\sum_{\nu lm}\frac{|\nu lm\rangle\langle 
    \nu lm|}{2 \nu^2} +2 \pi \int\ddn{3}{x}\rho(\vec x)a_s(k_{\vec x})|\vec x\rangle\langle\vec x|.
\ee

The eigenvalues $E_i$ and eigenstates $|\Psi_i\rangle = \sum_{\nu lm}c_{\nu lm}^{(i)}|\nu lm\rangle$ of this Hamiltonian are parameterized by the distribution of scatterer locations, $\rho(\vec x)$. We are interested in scatterers in a lattice configuration, and hence choose 
\be
\label{eq:scatdist}
\rho(\vec x) = \sum_{i=1}^{N_D}\delta^3(\vec x - \vec R_i),
\ee
\end{subequations}
where the scatterer positions $\vec R_i= d\sum_{j=1}^D n_{ij}\hat e_j$ are located at lattice positions described by unit vectors $\hat e_j$, the lattice spacing $d$, and a set of $D\times N_D$ integers $n_{ij}$. By excluding some values $i$ we can implement partial filling, defined by the fill factor $F$.

Eq.~\ref{eq:ham} relies on two more approximations: the scattering length is energy-independent and the basis is truncated to only a single $\nu$-manifold. We demonstrate in Sec. \ref{sec:homogeneous} that these approximations are increasingly accurate at high $\nu$.
They have only minor quantitative effects on the main conclusions of our study but allow us to obtain analytical formulas and clear scaling behavior. 
Since we consider only a single $\nu$-manifold, in the following discussion we will set  $-\frac{1}{2\nu^2}$ to zero. Hence,
the spectrum of Eq.~\ref{eq:ham} with the distribution from Eq.~\ref{eq:scatdist} is obtained by diagonalizing
%the pseudopotential operator, which has the matrix representation
\be
\label{pertmatrix}
V_{lm,l'm'} = 2\pi a_s\sum_{i = 1}^{N_D}\langle \nu lm|\vec R_i\rangle\langle \vec R_i|\nu l'm'\rangle.
\ee

\subsection{Scatterer induced properties in $D$-dimensional lattices}
\label{sec:latticeproperties}
 The properties of the Rydberg Composite, defined by the Hamiltonian in Eq.~\ref{eq:ham}, depend on the properties of both the unperturbed electronic states, $|\nu lm\rangle$, and the scatterer distribution, $\rho(\vec x)$. The Rydberg atom's size, density of states, and wavelength are determined by its principal quantum number $\nu$, while $\rho(\vec x)$ depends on the desired lattice geometry, lattice spacing, and filling realization.  In this section we delineate the important quantities  for one, two, and three-dimensional scatterer configurations. 

Although the spatial scale of the lattice greatly exceeds the size of the Rydberg wave function, not all scatterers perturb the Rydberg states since the electron-atom interaction is highly localized. Its strength, within the Fermi approximation, is determined by the electronic density directly at the scatterer position. The Rydberg volume is finite with a radius $r_0(l) \approx a_l\nu^2$, where $a_l$ decreases from $a_l\approx 2$ for the $l=0$ state to $a_l\approx 1$ for the $l = \nu-1$ states. Numerically, we consider scatterers inside a radius $r=a\nu^2$ with  $a>2$ to guarantee that all wave function amplitudes are exponentially small, and hence contribute no energy shift, at the boundary of this volume. The number $N_D$ of relevant scatterers is then determined by the volume $V_D$ of the intersection of the Rydberg wave function and the lattice. In this way, even for an infinite lattice, we can truncate its effect to only that caused by the $N_D$ individual scattering potentials in Eq.~\ref{eq:ham}.

A $1$D array of scatterers corresponds to a chain lattice, see Fig.~\ref{fig:introschematic}a.  The relevant $1$D volume is $V_{1}=2a\nu^2$, and $N_{1} = 2a\nu^2/d$ scatterers lie within this volume. The corresponding volume for a $2$D lattice is the area of the projection of the Rydberg volume into the plane, $V_{2} = \pi a^2\nu^4$, and hence the number of scatterers is $N_{2} = \frac{\pi a^2\nu^4}{d^2}$.  In $3$D we consider a cubic lattice of scatterers, and so the relevant volume is the entire Rydberg volume,  $V_{3} = \frac{4}{3}\pi a^3\nu^6$, containing  $N_{3} = \frac{4\pi a^3\nu^6}{3d^3}$ scatterers. The $N_D$ values given here are valid only in the $N_D\gg 1$ limit where edge effects due to the incommensurate spherical and cartesian geometries are negligible.

The scatterer configuration also influences how many of the degenerate states of the Rydberg manifold are shifted. As a general rule, each scatterer splits away one state from the degenerate manifold until the geometry induced limit $B_{D}$ is reached. In a generic $3$D scatterer array this limit is given by all states of the manifold, $B_{3}=\nu^2$, while $B_D < \nu^2$ in $1$D and $2$D. To determine $B_D$ for each case we select a convenient quantization axis and identify the  Rydberg states not affected by the delta-function potential (Eq.~\ref{eq:fermi}). In $1$D, we set the quantization axis parallel to the linear chain of scatterers. When $\vec r \to R\hat z$, most angular wave functions vanish on the quantization axis since
\be
Y_{lm}(\theta = 0,\phi) = \sqrt{\frac{2l+1}{4\pi}}\delta_{m0}.
\ee 
Only $m=0$ states experience a shift, and hence $B_{1}=\nu$ is the total number of $m=0$ states. For the $2$D case we set the quantization axis normal to the plane and evaluate the angular wave functions at $\theta = \pi/2$. The Legendre polynomials with argument $\cos(\pi/2) = 0$ are
\be
\label{eq:plm}
P_l^m(0)=
\begin{cases} 
(-1)^{(l+m)/2}\frac{(l+m-1)!!}{(l-m)!!} & l+m=\text{even} \\
     0 & l + m = \text{odd}\,.
   \end{cases}
   \ee
 The plane is transparent to the Rydberg states possessing a node in the plane. Therefore,
\be\label{eq:Tn}
B_{2}=\frac{\nu(\nu+1)}{2}\,.
\ee
With the help of $B_{D}$ we can define a third quantity, the characteristic lattice spacing $d_D$ such that $N_D\approx B_D$ for that geometry. This spacing heralds the onset of the density shift regime where additional scatterers cannot split away new states since the $\nu$-manifold is saturated. They instead contribute linearly to a mean-field energy shift, consistent with the conclusion drawn from the original applications of Fermi's pseudopotential \cite{AmaldiSegre,Fuchtbauer2}: the mean-field effect of the interaction of the Rydberg electron with the scatterers is an energy shift proportional to the electron-atom scattering length and to the scatterer density. The values for this characteristic length, along with the other values $V_D$, $N_D$, and $B_D$, are given in table \ref{tab:dimensiontable}. From these characteristic properties we can assess the behavior and crude scaling with $\nu$ of the Rydberg Composite for a given scatterer geometry. Notice that for $D\le 2$ the critical lattice spacing is linear in $\nu$, but follows $\nu^{4/3}$  for the $3$D case. 

   \begin{table}[t]
  %\textbf{Rydberg Composite Scaling Properties}\\
%\resizebox{\columnwidth}{!}{
\begin{tabular}{ p{4.5cm}| c c c c c  }
\hline\hline
\text{Dimension} ($D$) & 1 &\quad & 2 &\quad& 3 \\
\hline
& & & & &\\
 \text{Effective lattice volume} ($V_{D}$)\quad & $2a\nu^2$ &\quad & $\pi a^2\nu^4$ &\quad&$\frac{4}{3}\pi a^3\nu^6$\\
  & & & & &\\
\text{Number of scatterers} ($N_D$) \quad& $\frac{2a}{d}\nu^2$ &\quad & $\pi\nu^4\left(\frac{a}{d}\right)^2$ &\quad& $\frac{4}{3}\pi \nu^6\left(\frac{a}{d}\right)^3$\\
   & & & & &\\
\text{Number of shifted states} ($B_D$)\quad & $\nu$ &\quad& $\frac{\nu(\nu+1)}{2}$ &\quad &$\nu^2$\\
  & & & & &\\
Maximum lattice spacing ($d_D$) such that $N_D=B_D$ \quad &$\frac{a\nu}{2}$ &\quad & $\sqrt{2\pi}a\nu$ &\quad&$\sqrt{\frac{4\pi}{3}}a \nu^{4/3}$\\
 \hline\hline
\end{tabular}
   \caption{Rydberg Composite scaling properties}
 \label{tab:dimensiontable}
\end{table}

This analysis suggests that for sufficiently large number of scatterers $N_D$ we will obtain $B_D$ non-zero eigenvalues upon diagonalizing $H$ within a $\nu$-manifold, and as a function of decreasing $d$ these eigenvalues will grow (on average) linearly with the number of scatterers. In order to remove this asymptotic shift we normalize the total energy shift by $N_D$. Furthermore we measure energies in units of $(2\pi |a_s|)^{-1}$ in order to remove the numerical prefactor from the potential matrix (Eq.~\ref{pertmatrix}), and hence eliminate the material-dependent value of the scattering length from our calculated energy shifts. Finally, since $N_D$ depends on the  arbitrary (provided it is sufficiently large) choice of $a$, we scale the energy shifts of the $D$-dimensional lattice by $a^D$ to eliminate this scale choice. Of course, in the limit $a\to\infty$, this choice removes all dependence on $a$ and we can  report scaled energies $\tilde{E}$ defined in terms of the un-scaled eigenvalues $E$ via
\be
\label{eq:energies}
\tilde{E} =\left(\frac{d}{\nu^2}\right)^D\frac{E}{2\pi |a_s|\tilde{V}_D},
\ee
 where  $\tilde{V}_D$ is the volume of a  $D$-dimensional sphere with unity radius.
We now investigate the behavior and properties of the Rydberg Composite by computing its spectrum for each geometry.

\section{Phenomonology of the Rydberg Composite}
\label{sec:phenomenology}
The spectrum of a Rydberg atom immersed in a structured neutral medium depends both on the Rydberg principle quantum number and on the different realizations of the lattice. We parameterize the latter by its filling factor $F$, the percentage of filled lattice sites, and by the lattice spacing $d$. We focus first on unity filling factor so that we can introduce the essential quantities useful in characterizing the Composite's properties. In section  \ref{sec:evolution} we will remove this restriction and study  fractional filling. 

We first study the density of states (DoS). It reveals more about the global spectral properties  than individual energy levels, and provides a useful guide to regions of interest to focus on in finer detail. In a second step, guided by the features seen in these DoS, we study the wave functions corresponding to various paradigmatic states.  The structure present in these wave functions provides additional investigative tools to understand the spectra. Since the $2$D monolayer leads to the richest structure in the dense lattice limit, we focus on that geometry.

%A Rydberg atom immersed in a perturbing mono-layer
\subsection{Density of States}
\label{sec:dos}
\begin{figure}[t]
    \centering
    \includegraphics[width=\columnwidth]{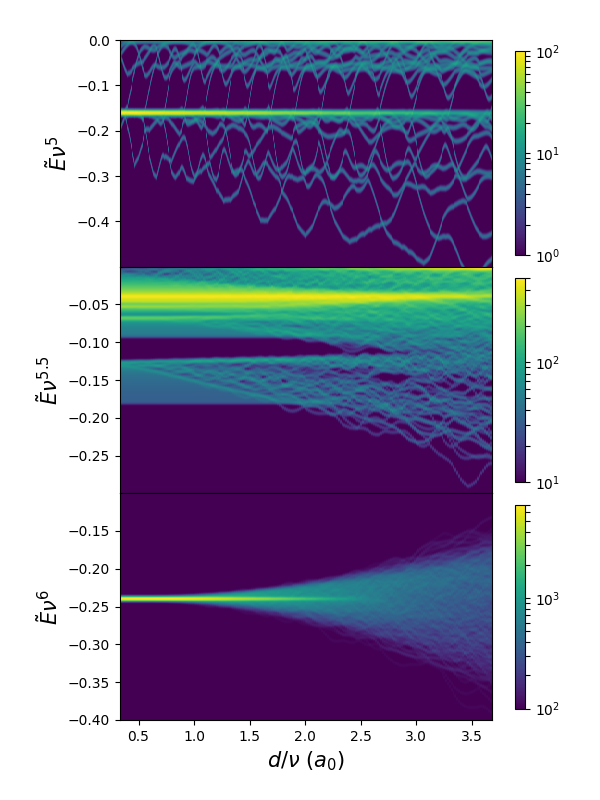}
    \caption{ Density of states (DoS) for a $\nu=30$ Composite in $1$D (a), $2$D (b), and $3$D (c) as a function of scaled lattice spacing $d/\nu$. The scaled energy units (Eq.~\ref{eq:energies}) show that in all three cases the $N_D\to\infty$ form of the DoS  becomes constant. 
    %The energy scale is set to zero at the bare Rydberg energy $E=-\frac{1}{2\nu^2}$.
    The color is the DoS, ($\frac{\text{d}N}{\text{d}\tilde{E}}$).% For $1$D and $3$D the spectrum becomes degenerate again in the dense limit, while the $2$D spectrum reveals a rich band-type structure. 
    }
    \label{fig:alldDoS}
\end{figure}

We show DoS in Fig.~\ref{fig:alldDoS} for the lattice geometries depicted in Fig.~\ref{fig:introschematic}.  We observe that all $B_D$ eigenstates converge to constant limits for $d\to0$, as anticipated. Intriguingly, we find that the asymptotic value differs remarkably across the three geometries. For the $1$D and $3$D scatterer geometries the shifted eigenenergies become degenerate again as $N_D\to\infty$, albeit at a large overall energy shift relative to the zero-scatterer degenerate manifold. In contrast, eigenenergies in the $2$D geometry remain non-degenerate even in the infinite density limit, instead developing three main features (see also the spectrum Fig.~\ref{fig:2d}): a nearly continuous and quasi-uniformly spaced distribution of energy levels within a few ``bands'', the formation of a large ``band gap'' that persists even up to relatively large lattice constants, and the formation of a large peak in the DoS in the upper part of the  spectrum. 

In all three lattice geometries, as $d$ increases the DoS becomes challenging to interpret due to the increasing number of non-degenerate energies. In general, the spectrum diffuses.  In $1$D, the degenerate band is depleted as individual states split discretely away with increasing $d$. This process does not occur symmetrically with respect to the degenerate band. In $3$D, all states begin to split apart at approximately the same value of $d$ and the perturbed band dissipates far more rapidly than in $1$D; this process also occurs symmetrically about the homogeneous energy asymptote. In $2$D the states are 
not degenerate in the $d\to 0$ limit. For increasing $d$, states
%non-degenerate as $d\to 0$, giving a smooth DoS, and so as $d$ increases we see that the DoS begins to become dependent on $d$. States 
higher in the energy band begin to disperse linearly in $d$, revealing a clear energy dependent transition between the indistinguishable ($d\approx0$) and distinguishable scatterer case.
%between the two regimes that is approximately a linear function of $d$.
In all three geometries, oscillations in the energy levels mimic the oscillatory nature of the Rydberg wave function, which is imposed quite directly onto the energy levels via the contact potential.  The ``spaghetti'' nature of the energy levels in the large $d$ regime reveals the presence of both, real and avoided level crossings if $d$ is taken as an adiabatic parameter. Real crossings are possible since the electronic states mirror the lattice symmetry, and therefore can be grouped according to the irreducible representations of the nuclear point group for that lattice. We have confirmed that, in the $2$D lattice case, the DoS can be computed independently for each of the five irreducible representations of the $C_{4v}$ point group, following the description of Refs. \cite{Rost2006,EilesJPB}. 
%\red{As the two-dimensional E irreducible representation contains by far the majority of the states, this two-fold structure corresponds  to the two groups of levels with lower and higher energies, clearly distinguishable in  Fig.~\ref{fig:alldDoS}.}
This is discussed in more detail in Appendix \ref{app:symorbitals}. Finally, as $d$ grows further, the DoS (not pictured) collapses gradually back into a highly degenerate peak at zero energy as the number of scatterers falls below $B_D$.

\subsection{Wave function characteristics}
\label{sec:wavefunctions}
We now present a representative sample of the wave functions giving rise to these DoS in $1$D and $2$D, which are  particularly amenable to this treatment since all relevant information can be gleaned and easily visualized with three-dimensional contour plots ($1$D) or the $z=0$ slice through the electron density ($2$D). From these wave functions we begin to see the underlying structure of the perturbed system and how it might lead to  the emergence of the structured, non-degenerate bands in $2$D band structure rather than  a single, fully degenerate band in $1$D and $3$D. Although our focus now is descriptive, merely commenting on the appearance and classification of these wave functions, we will use these observations in the following section to develop quantitatively accurate approximations which lead to a full interpretation of the Rydberg Composite's properties.

\begin{figure}[t]
    \centering
    \vspace{-50pt}
    \includegraphics[width=0.8\columnwidth]{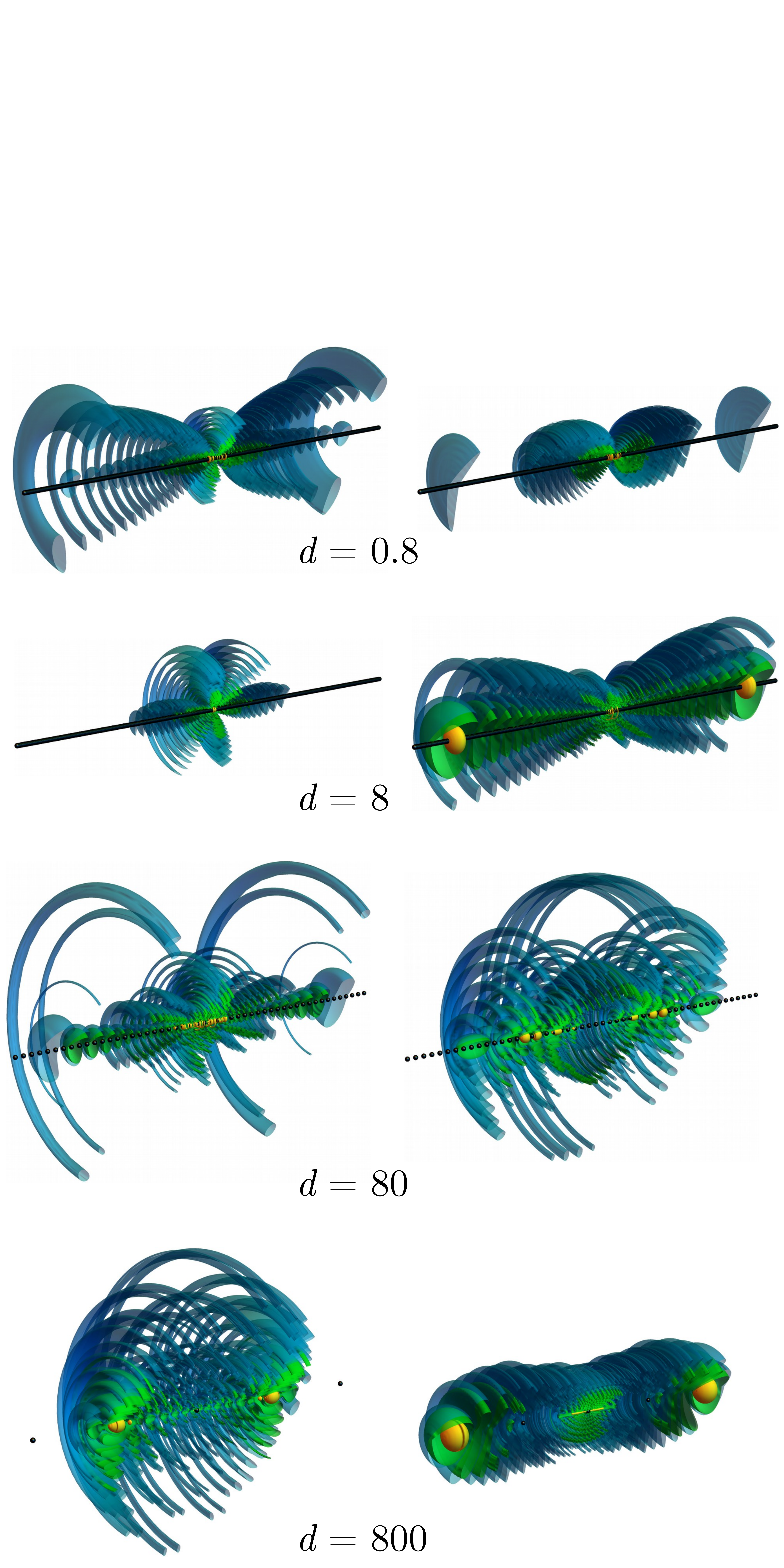}
    \caption{Electronic densities of a 1D chain of scatterers for $\nu=30$. The yellow, green, and blue surfaces correspond to contours at wave function densities spanning factors of 10. The full three-dimensional surfaces are cut away in front to reveal the interior structure.  The left column gives the density for the most deeply perturbed state, while the right column gives the density just above the degenerate band. }
    \label{fig:1dwfs}
\end{figure}

\begin{figure*}[t]
%\centering
\hspace*{-1cm}
\includegraphics[width=\textwidth]{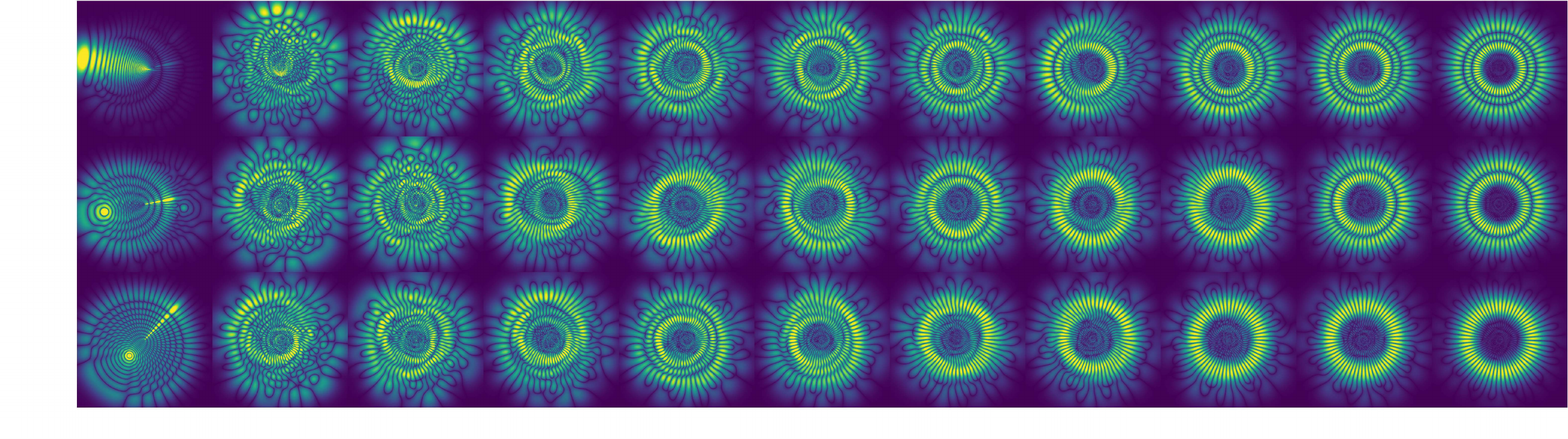}\\%\hspace{-2cm}
%a; d=[20,54,60,80,100,1000] e[1-3], b; d=[20,54,60,70,80,90,100,200,500,1000] e[1-5],c; d=[20,54,60,70,80,90,100,200,300,500,1000] e[1,3,5]
%\centering
\hspace*{-1cm}
\includegraphics[width=\textwidth]{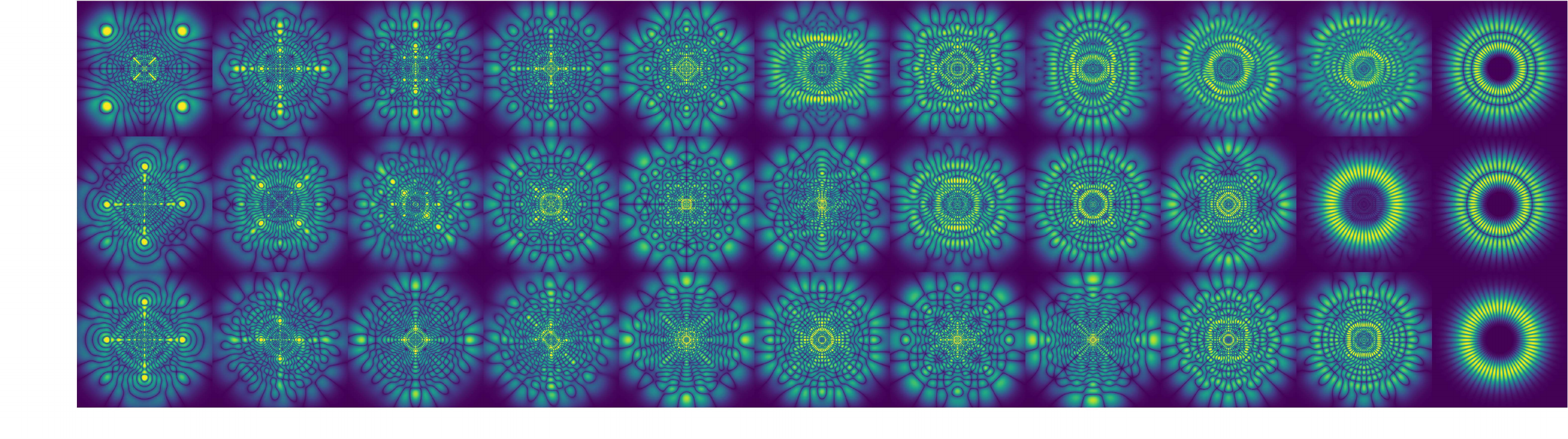}

%\red{I recommend having both panels start with the $d\to0$ limit, and also flipping a and b. Iprefer this as it read like a book }
\caption{The electron density  $|\Psi(\vec r)|^{\frac{1}{2}}$ of selected eigenstates of a $\nu =30$ $2$D-Composite. Both panels display from bottom to top the first, third, and fifth eigenstates.
(Top) Probability densities for a dense lattice ($d=20$) increasing in fill factor from left to right in steps of 0.1 excluding the first panel where only five scatterers are present. 
%The y-axis is ordered in increasing energy showing the 1st, 3rd, 5th eigen energy respectively
(Bottom) Probability densities for a full lattice with $d=\{1000, 500, 300, 200, 100, 90, 80, 70, 60, 54, 20\}$, respectively.
%and increasing in energy up the way (1st, 3rd, 5th eigenenergy respectively).
}

%d=(20,54,60,70,80,90,100,200,300,500,1000)
%\includegraphics[width=\textwidth]{fig1b.pdf}
   
    \label{fig:filling_up}
\end{figure*}
\subsubsection{1D lattice wave functions}
We present, for four different lattice spacings, two representative wave functions for the  $\nu=30$ $1$D Rydberg Composite. On the left we show the state with the largest energy shift, while on the right we choose a state slightly higher in energy than the degenerate band limit, i.e. one of the states visible in Fig.~\ref{fig:alldDoS}a just above the middle band.  These wave functions visually forge the connection between Rydberg Composites and ``trilobite'' molecules \cite{Rost2006,Rost2009,EilesJPB,PhysRevA.99.022506}. At large $d$, shown in the bottom row, the  wave function is a mixture of many $l$ states, leading to strong localization on scatterer positions. In scenarios such as this where scatterers are separated by distances greatly exceeding $d_D$, the wave function tends to localize on only a subset of the scatterers and effectively ignore the rest. In this way it maximizes the overlap between the Rydberg electron and the lattice, and ensures orthogonal wave functions. As $d$ decreases these states eventually begin to resemble the hydrogenic basis states and localize less severely on a symmetry-imposed collection of scatterers, as the Rydberg wave function increasingly cannot distinguish scatterers lying closer together than its wave length.  Unfortunately, these wave functions do not as yet reveal with any clarity why the infinite density limit of this $1$D-Composite is again an energetically degenerate system. A key reason for this uncertainty is, in fact, their degeneracy: degenerate eigenstates obtained via a numerical diagonalization will in general be arbitrary superpositions of the, linearly independent, states. It is thus impossible to identify any possible good quantum numbers or selection rules from these wave functions without investigating some other observable. In principle, this could be done by applying a magnetic field to break apart the degeneracy at large scatterer density. For our present purposes we can turn instead to the $2$D-Composite, which is fundamentally non-degenerate in this limit and may reveal through its wave functions the underlying structure of the $1$D case.

\subsubsection{$2$D lattice wave functions}
Since only the electronic density in the $z=0$ plane contributes to the energy shifts, it suffices to examine $|\Psi(x,y,0)|$ for the $2$D-Composite. We first consider $2$D-Composite wave functions with a fully filled $2$D lattice and vary the lattice constant. In Fig.~\ref{fig:filling_up}(Top) we show the wave functions corresponding to the first three odd-numbered eigenenergies starting from the lowest one. For large enough $d$ the electron density obeys one of the discrete symmetries permissible by the $C_{4v}$ point group, and partially localizes on only a subset of the available scatterers. The behavior of this localization and its effect on the energy level structure likely warrants future study. As $d$ shrinks further, the electron density evolves into a distinctly circular shape. By the lowest $d$ shown ($d=20$), these three eigenstates have seemingly converged into ``circular'' states. By a strict definition, a circular Rydberg state has $l = m = \nu-1$; here we employ a broader definition meaning a state with high $l$ and $|m|$, but with only a small difference $l-|m|$. 
The second and forth eigenstates identically resemble the first and third, respectively, showing that the $\pm m$ states are equivalent and degenerate in this limit
%The even-numbered eigenstates not pictured resemble in this small $d$ limit the first and third eigenstates, confirming that both states with $\pm m$ are equivalent in this limit.

To confirm that these eigenstates do not arise due to some coincidence in the symmetry-adapted wave functions or fortuitous overlap with the lattice grid, we next consider a lattice with a small lattice constant $d\ll d_D$ but with varying fill factor $F$. At extremely low $F$ (first column of Fig.~\ref{fig:filling_up}(bottom)), having only a very few scatterers, we see that the few non-degenerate eigenstates are basically independent trilobite dimers between the Rydberg core and each individual scatterer. As $F$ increases the number of scatterers increases rapidly and the wave function becomes rather chaotic in appearance, exhibiting no clear structure.  In some instances it localizes asymmetrically about statistical fluctuations in the random scatterer distribution where small clusters form spontaneously.

As before, when the series progresses towards complete filling, the density resembles more and more a circular state, thus confirming that the appearance of such states depends more on the total density of scatterers relative to the number fluctuations caused by random fill factors than on the underlying lattice symmetry. Once fluctuations and correlations in the scatterer density are unresolved by the Rydberg wave function, any choice of random fill factor is essentially indistinguishable and the result from the $F=1$ case is reached.

\subsection{Role of wave function character on the $2$D spectrum}
 
\begin{figure}[t]
    \centering

        \includegraphics[width=\columnwidth]{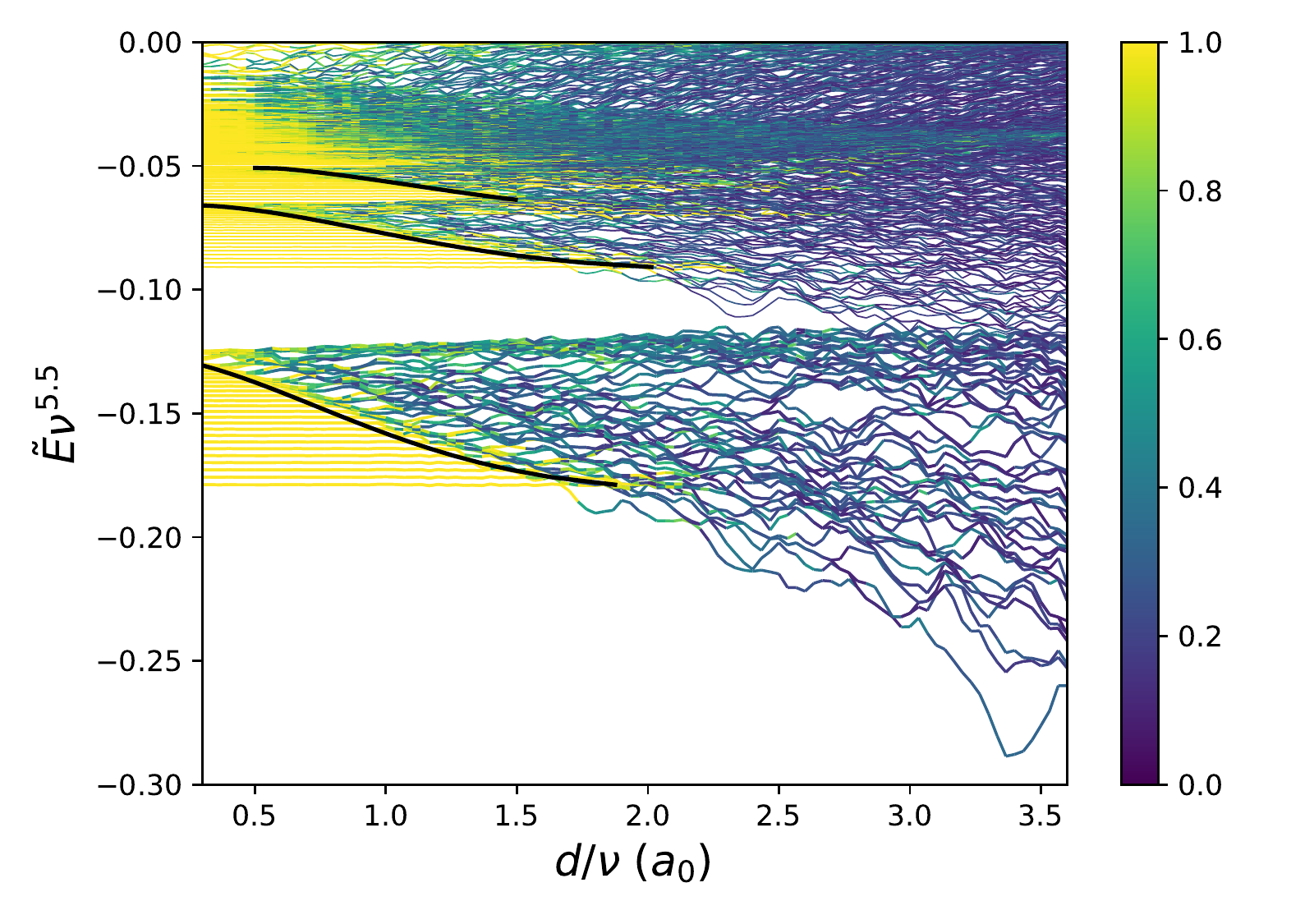}
    \caption{ $2$D-Composite spectra for a $\nu=30$ Rydberg Composite as a function of lattice spacing. The line color shows $P_m$ for each state as defined by Eq.~\ref{eq:IPR}. The black line marks the border between homogeneous and inhomogeneous regimes (See Eq. \ref{andrew}.).}
    %traces out $d_\mathrm{c}$ from Eq.~\ref{andrew}. }
    \label{fig:2d}
\end{figure}

Both ways of increasing the scatterer density described above lead to the following conclusion in the high density limit: the wave functions become increasingly circular in character, implying that they become approximate eigenstates of $\hat L_z$.  This is to be expected in the limit of a totally homogeneous lattice, where $H$ commutes with $\hat L_z$ due to the cylindrical symmetry. Of greater interest is the fact that the energies of these states are also apparently sorted by the level of circularity, as states with the most circular character fall to the bottom of the  energy bands.  A useful diagnostic to analyze the evolution of the wave functions as $d$ or $F$ changes is the participation ratio of $m$ states,  
\be
\label{eq:IPR}
P_m=\sum_{m'=\pm m}\left( \sum_l |c_{lm'}|^2\right)^2,
\ee 
which ranges from $1$ for a state proportional to $\delta_{mm'}$ to $1/\nu$ for a state mixed uniformly among $m$ sub-levels. In Fig.~\ref{fig:2d} we show the $2$D-Composite's eigenspectrum as a function of $d$. Although this conveys very similar information as the DoS plot in Fig.~\ref{fig:alldDoS}, coloring the eigenstates by $P_m$ reveals additional structure in these energy levels that can be linked to the wave function. The $P_m$ distinguishes many self-similar and repeating substructures that were not evident in Fig.~\ref{fig:alldDoS}b. Several ``bands'' of states with a similar functional dependence on $d$ and pattern of $P_m$ are visible, separated by the large energy gap that was clear in the DoS as well. These bands become indistinguishable towards high energy and converge into the region of high degeneracy seen in Fig.~\ref{fig:alldDoS}. The clear transition between states with $P_m\approx 1$, which have $m$ as a good quantum number, and those which are strongly mixed helps differentiate these bands even when they start to overlap. This transition is well predicted by a critical lattice spacing ($d_\mathrm{c}$) defined in Sec. \ref{sec:evolution} and shown for the first three bands in Fig.~\ref{fig:2d} as black curves.
Figure  \ref{fig:2d} shows that the trend towards circular states in the few eigenstates presented in Fig.~\ref{fig:filling_up} is emblematic of a more general behavior: in the $d/\nu \ll 1$ limit $H$ commutes with  $\hat L_z$ and hence all wave functions have $m$ as a good quantum number, of which the circular states are a small subset. From the wave functions in Fig.~\ref{fig:filling_up} we see also that the  lowest states of the energy bands approximately conserve $l$ as well, with maximal or nearly maximal values of both quantum numbers.

We will devote much of the reminder of this paper to the $2$D-Composite, since embedding the Rydberg excitation in a planar environment  constitutes a new scheme in ultracold Rydberg physics with a rich and intricate behavior. In particular, we will explicate the physics underlying these (so far phenomenological) observations: the formation of energy bands separated by a single dominant band gap and the relatively simple character of the underlying wave functions. We elucidate the link between this structure and the structure underlying the Hamiltonian matrix. The mathematical tools used for this task also enable us to understand why all states become degenerate in a $1$D and $3$D homogeneous environment but not in a planar one. Furthermore, these tools prove useful as a launching point for our later investigation of disorder in dilute lattices with random filling and lattices with larger $d$.

\section{Rydberg Composite Properties in the Homogeneous density limit}
\label{sec:homogeneous}
Our investigation of the $2$D-Composite spectrum starts with the observation that, below a certain lattice constant $d<d_\mathrm{c}$, the Rydberg wave function can no longer resolve individual scatterers. The lattice then appears homogeneous, and the phenomenology of previous sections has shown that the spectrum becomes constant.  In section \ref{sec:evolution} we will clarify this coarse-graining concept further for the disordered scenario. In the present section, we will take it as fact that this coarse-graining is physically relevant and use it to approximate the discrete lattice of scatterers with a continuous plane of homogeneous density.  In this way we  characterize the system's properties for the $d\ll \nu$ region  of Figs. \ref{fig:alldDoS} and \ref{fig:2d}, which is then crucial to properly situate our analysis for intermediate cases with  $d>d_\mathrm{c}$ or  $F <1$.  

\subsection{2D Monolayer: emergence of a band structure}
\label{sec:monolayer}
The replacement of the discrete lattice with a homogeneous distribution coincides mathematically with the replacement of the summation  in Eq.~\ref{pertmatrix} with an integral. In the scaled energy units this replacement must include also a factor $V_2^{-1}$, and the matrix elements become
\begin{equation}
\label{eq:matrixelement2d}
\begin{split}
\lim_{d\to 0} \tilde{V}_{lm,l'm'}= \frac{\int\Psi_{\nu lm}^*(R,\frac{\pi}{2},\varphi)\Psi_{\nu l'm'}(R,\frac{\pi}{2},\varphi)\dd{\mathcal{A}}}{V_{2}}\,
\end{split}
\end{equation}
with integration over the entire plane. Using the spherical coordinate representation of these wave functions, the three contributions to the matrix elements  are given by the product of an integral over $\varphi$,
\be
\int_{0}^{2\pi}e^{i(m-m')\varphi}\dd{\varphi} = \delta_{mm'},
\ee
a radial overlap integral,
\be
\mathcal{R}_{\nu l,\nu l'}^{(j)}=\int_0^\infty \frac{u_{\nu l}(R)u_{\nu l'}(R)}{R^j}\dd{R},
\ee and the projection of the spherical harmonics into the plane, $\mathcal{P}_{lm,l'm}=N_{lm}P_l^m(0)N_{l'm'}P_{l'}^{m'}(0)$, where  
\begin{align}
 N_{lm}&=\sqrt{\left(l + \frac{1}{2}\right)\frac{(l-m)!}{(l+m)!}}
\end{align}
and $P_l^m(\cos\theta)$ was given in Eq.~\ref{eq:plm}. As expected, integration over $\varphi$ imposes a block-diagonal structure in $m$ on this matrix, since the homogeneous scatterer limit is isotropic.  Eq.~\ref{eq:matrixelement2d} therefore yields
\be
\label{junk11}
 \lim_{d\to 0}\tilde{V}_{lm,l'm'} = \frac{\delta_{mm'}}{\pi \nu ^4}\mathcal{P}_{lm,l'm}\mathcal{R}_{\nu l,\nu l'}^{(1)}\,.
\ee

\begin{figure}[t]
    \centering
    \includegraphics[width=\columnwidth]{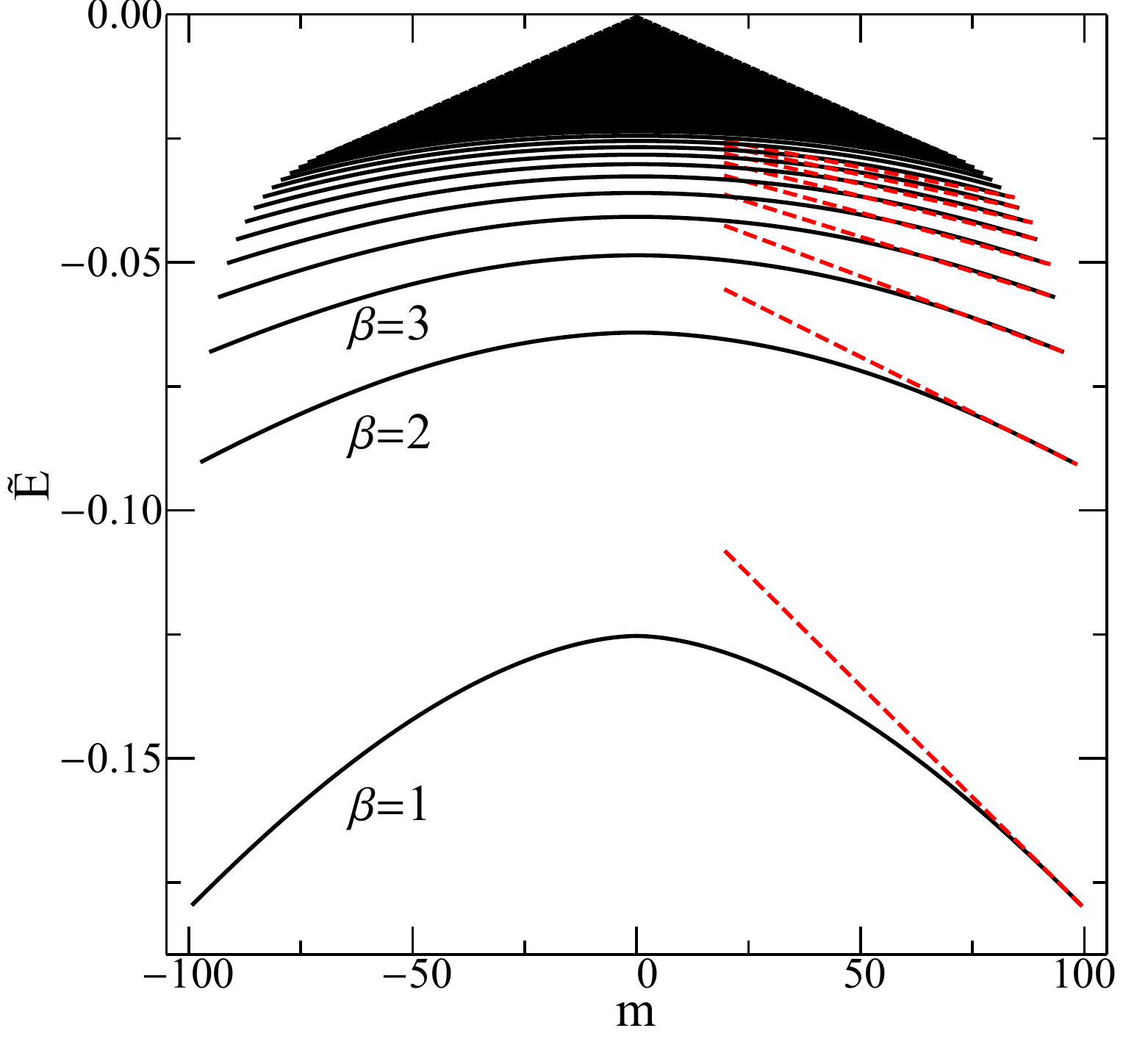}
    \caption{The Rydberg Composite spectrum displayed as a dispersion relation, emphasizing the formation of energy bands. The black curves show $\widetilde{E}(m/\nu)$ for $\nu = 100$.  The dashed red lines are the asymptotic linearization of Eq.~\ref{wings}. }
    \label{fig:bandstates}
\end{figure}

\noindent Fig.~\ref{fig:bandstates} displays the eigenvalues of this block-diagonal matrix, plotted as a function of $m$ to  emphasize the parallels with a band structure. We see that the resulting eigenvalues can, by connecting the ranked eigenvalues across $m$ values, be sorted into energy bands which are linear in the wings at high $|m|$ and quartic near $|m|=0$. We label these with a band index $\beta$,  thereby characterizing each eigenenergy by a $(\beta,m)$ label. As $\beta$ increases the wings of upper bands begin to overlap the flat low-$|m|$ regions, and we find in this overlapping region that each band begins along the essentially continuous line $\tilde{E}=-\frac{m}{\sqrt{2\pi \nu^3}}$.

\begin{figure}[t]
    \centering
    \includegraphics[width=\columnwidth]{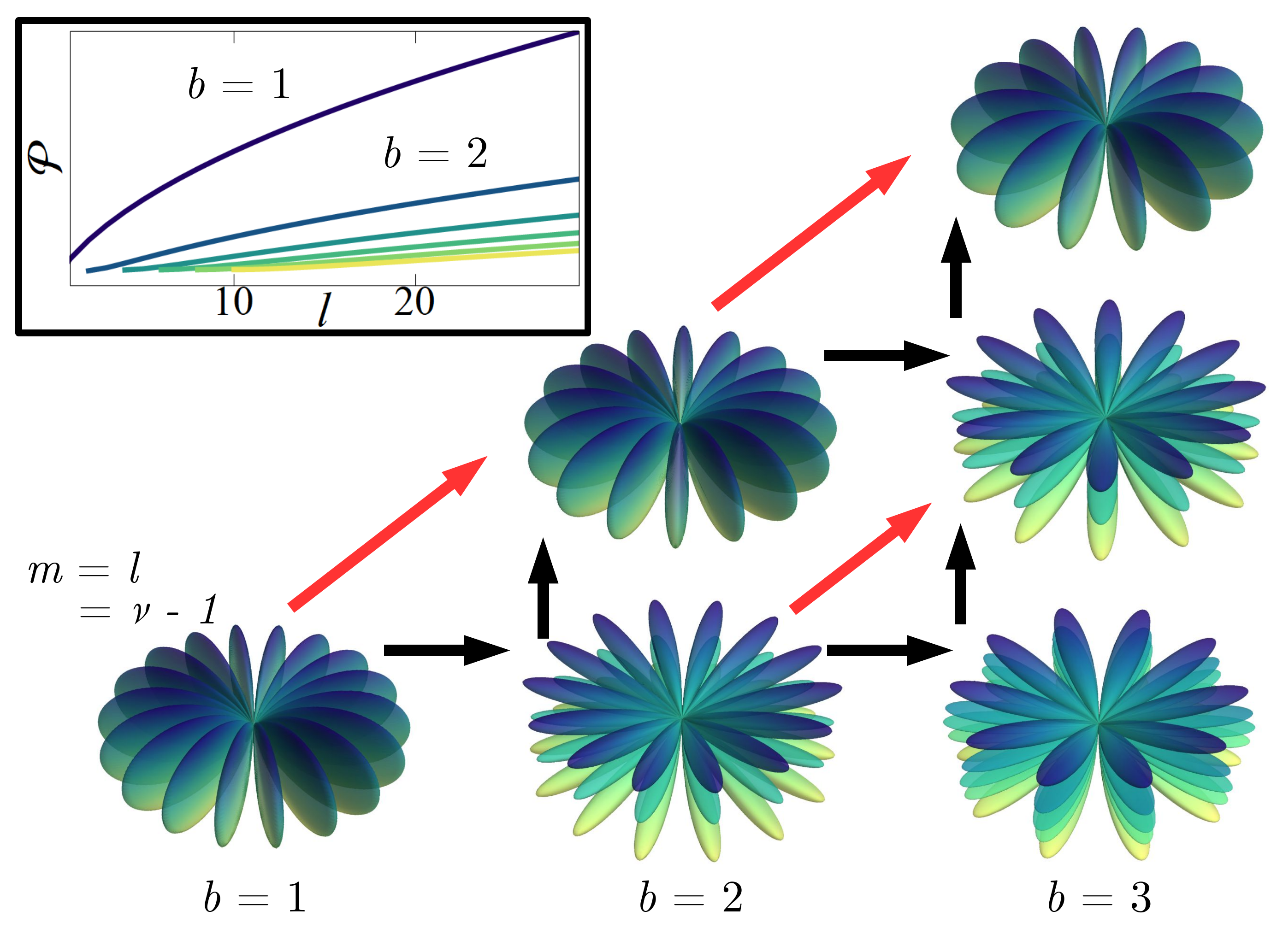}
    \caption{Dependence of the band structure of a 2D Rydberg Composite on the  projection of $Y_{lm}(\theta,\varphi)$ into the plane. Shown are a few spherical harmonics, starting in the bottom left corner with the maximally circular state. Horizontal black (vertical black) arrows represent a decrease in $m$ ($l$). States in the same column  therefore are from the same block-diagonal. Red diagonal arrows represent increases in $k=\nu-l$, starting from $k=1$ in the bottom row. The inset shows the projection $\mathcal{P}$  as a function of $l$ for $\nu = 30$ for several $b$-bands. Notice how much larger the drop in energy is between bands compared to the gaps between $k$-states within a band. }
    \label{fig:sphharmos}
\end{figure}

To understand the formation of these bands as well as to obtain analytic results for the eigenspectrum, we study the matrix elements of each $m$-level block of the Hamiltonian. As one might surmise from the wave function study in the previous section, these sub-blocks appear to be diagonal-dominated for moderate to high $l$ and $m$. This can stem from two influences. First, the off-diagonal couplings
%can be small compared to the diagonal matrix elements, and indeed they do 
tend to be around one order of magnitude smaller than the diagonal matrix elements. This is because, although the radial overlap integral does not have a rapid dependence on $l$, the $u_{\nu l-2}(R)$ wave function has a node nearly at the maximum antinode of the $u_{\nu l}(R)$ function, and hence the integrand of $\mathcal{R}_{\nu l,\nu l'}^{(1)}$ is small compared to $\mathcal{R}_{\nu l,\nu l}^{(1)}$. An additional and more critical contribution stems from the spherical harmonic projections onto the plane, which are illustrated with surface contour plots in  Fig.~\ref{fig:sphharmos} for these circular states. Each column of this figure contains all the states within a single  $m$-block, starting with $m=\nu-1 = l$ on the left and decreasing by $2$ (because of the selection rule of Eq.~\ref{eq:plm}) with each step to the right. The orbital angular momentum $l$ decreases by 2 with each vertical step up from its maximal value, $l = \nu - 1$, in the bottom row. Since $l$ cannot be less than $|m|$ the size of each sub block increases with decreasing $m$.  There is no coupling between columns, which come from different $m$-blocks. Clearly, the states within each column have dramatically different overlap with the $z=0$ plane since each drop in $l$ pushes an additional lobe out of the plane. These contribute nothing to the total energy shift and are essentially ``wasted'' probability.  In contrast, along the diagonals marked by red the wave functions are nearly identical. Moving up the diagonal swaps an angular lobe into a (not pictured)  radial lobe, which has a negligible impact on the overlap with the plane compared to the loss of pushing an entire set of lobes out of the plane. Within each $m$-block, therefore, the diagonal elements will have large energy separations, and hence effectively decouple. The resulting states that share similar qualities are spread over many blocks and correspond to the series along the diagonal in Fig.~\ref{fig:sphharmos}. We label elements in these series with the values $b=1\dots \nu-1$ and $k=1\dots \nu-b$. These numbers label the diagonals and the state in a given diagonal, respectively. Note that from the construction in Fig. \ref{fig:sphharmos} we have $l=2 b-2+m$ and $k=\nu-l$.  The Fig.~\ref{fig:sphharmos} inset displays the overlap $\mathcal{P}_{lm,lm}$ for various $b$ as a function of $l$. Each $b$-band smoothly changes with $l$, but the difference between $b$-bands is large, especially so for the lowest $b$ values. 

The appearance of bands in the eigenspectrum (see Fig \ref{fig:bandstates}) is essentially a consequence of the approximate $b$-bands of similar states described in Fig  \ref{fig:sphharmos}.
The identity $b=\beta$ holds exactly when the $m$-block  matrices are diagonal, i.e. in the asymptotic wings of these bands where the dispersion becomes approximately linear (see Fig. \ref{fig:bandstates}) and where $l\approx m \approx \nu$. Since in this limit the eigenenergies are obtained analytically and the real bands $\beta$ coincide with the approximate bands $b$, we focus now on the behavior of these linear wings.  If we consider only the diagonal matrix elements of Eq. \ref{junk11},
\begin{align}
\lim_{d\to 0}\tilde{V}_{lm,lm} &=\frac{1}{\pi \nu^4}\left[N_{lm}P_l^m(0)\right]^2{R}_{\nu l,\nu l}^{(1)}\nonumber\\
%&=\frac{1}{A}\left(l + \frac{1}{2}\right)\frac{(l-m)!}{(l+m)!}\left(\frac{(l+m-1)!!}{(l-m)!!}\right)^2\frac{1}{n^2}\\
&=\frac{1}{\pi \nu^4}\frac{2l+1}{2^{2l+1}\nu ^2}\frac{(l-m)!(l+m)!}{\left[\left(\frac{l+m}{2}\right)!\left(\frac{l-m}{2}\right)!\right]^2},
\label{eq:test}
\end{align}
we obtain the energy levels for the $\sim 10$ lowest energy levels in each band of states to a few percent accuracy, confirming that the diagonal approximation is appropriate. More importantly, by switching to the band numbers $b$ and $k$ we can gain further intuition into the true bands, labeled by $\beta$, seen in Fig. \ref{fig:bandstates}. In the  high $\nu$ limit we obtain the linear dispersion relation
\be
\label{wings}
\tilde{E}^{2D}_{bk}\approx\frac{(8\nu +4(b-k)-1)\Gamma(b-1/2)}{8\pi^2\nu^{13/2}\Gamma(b)}.
\ee
In particular, band $b\approx \beta$ starts at the energy 
\be
\label{eq:band_start}
\tilde{E}^{2D}_{b1}=\frac{\Gamma(b-1/2)}{\Gamma(b)}\frac{1}{\pi^2 \nu^{11/2}}.
\ee
For  $b\ll\nu$ the energies scale as $\nu^{-11/2}$. In particular, the lowest energy lies at $\tilde E = \nu^{-11/2}\pi^{-3/2}$.  At higher $b$ we use the limiting form of the $\Gamma$ functions,
\be
\lim_{b\to \infty}\frac{\Gamma(b-1/2)}{\Gamma(b)} = \frac{1}{\sqrt{b}}
\ee
to obtain
\be
\lim_{b\to \infty}\tilde{E}^{2D}_{bk}=\frac{1}{\pi^2 \sqrt{b\nu^{11}}}.
\ee
The band's lower edges become more closely spaced in energy due to this $1/\sqrt{b}$ dependence.  The level spacing within  a band is given approximately by 
\be
\lim_{b\to \infty}\Delta_k = \frac{\Gamma(b-1/2)}{2\nu^{13/2}\pi^2\Gamma(b)}=\sqrt{\frac{1}{b}}\frac{1}{2\pi^2 \nu^{13/2}}.
\ee
Taking this width as approximately constant over an entire band and taking the number of states per band to be $\sim\nu$, we find that the width of each band is approximately
\be
\Delta \approx \frac{1}{2\pi^2}\frac{1}{\sqrt{b}\nu^{11/2}}.
\ee
On the other hand, the spacing between band minima is approximately
\be
\label{eq:spacings}
\Delta_b \approx  \frac{\text{d}}{\text{d}b}\frac{1}{\sqrt{b\nu^{11}}}\sim\frac{1}{\sqrt{b^3}\nu^{11/2}} .
\ee
Within this crude series of approximations we find that $\frac{\Delta_b}{\Delta}\propto \frac{1}{b}$. Due to this decreasing gap between bands relative to their own widths the bands begin to overlap, leading to the region of high energy density seen in Figs. \ref{fig:alldDoS}b and \ref{fig:2d}, and apparent in Fig.~\ref{fig:bandstates}. As the bands overlap with increasing $b$, the expression for the band minimum Eq.~\ref{eq:band_start} tends towards $\nu^{-6}$, as $b\sim \nu$. Apparently, as the energy-level structure transitions from the ``band'' type into this denser structure of many overlapping bands, the functional form of the energy scaling changes from $\nu^{-11/2}$ to $\nu^{-12/2}$.

\subsection{Scaling laws for scatterers in $D$-dimensions}
\label{sec:scalingproperties}

\subsubsection{2D scatterers}

\begin{figure*}[t]
    \centering
    \includegraphics[width=\textwidth]{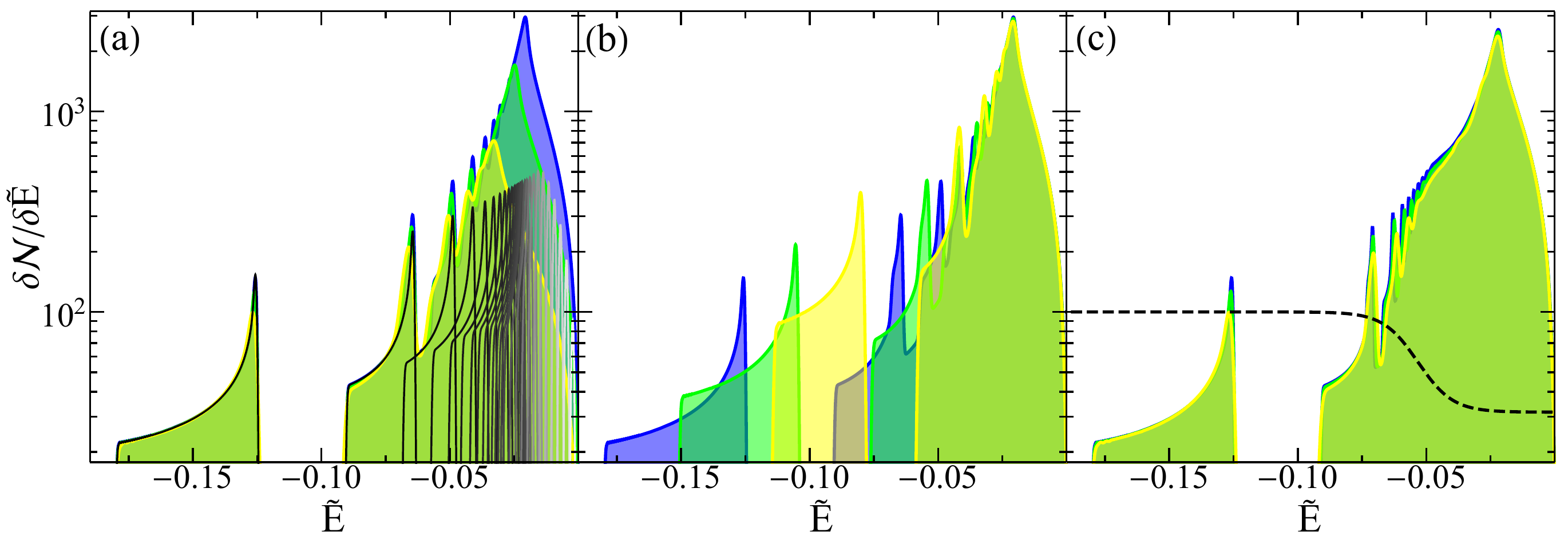}
    \caption{Density of states computed by 
    convolving the spectrum using a Gaussian distribution
    %binning the eigenenergies via Gaussian distributions 
    for three $\nu$ values: $40$ (yellow), $70$ (green), and $100$ (blue). (a): The energies are scaled with $\nu^{5.5}$, the ``band'' scaling. Individual band contributions for $\nu = 100$, obtained from the dispersion curves in Fig.~\ref{fig:bandstates}, are shown in grayscale.  (b) the energies are scaled with $\nu^{6}$, the ``overlap'' scaling. (c) The energies and densities of states are scaled via the ``universal'' scaling connecting the band and overlap regions. The interpolating $\tanh$ function connecting the two regimes is overlayed. 
    %$\sigma$ is chosen to be 0.01 for the Gaussian case, corresponding to a 
    The FWHM of the convolving Gaussian distribution is of approximately 0.0235. }
    \label{fig:DOS}
\end{figure*}

The preceding analysis showed that the energy spectrum of the $2$D-Composite exhibits two different scaling behaviors as a function of band number. The energies in the lower bands scale as $\nu^{-5.5}$, but they scale as $\nu^{-6}$ in the upper bands  due to a mixture of overlapping bands and deviations from the diagonal approximation for small values of $l$ and $m$. 

\subsubsection{1D scatterers}

Applying this same analysis to our $1$D and $3$D configurations leads quickly to the results that all states experience an identical energy shift. In $1$D, the  expression equivalent to Eq.~\ref{eq:matrixelement2d} is
\begin{align}
 \lim_{d\to 0}\tilde{V}_{lm,l'm'} 
    &=\frac{a}{V_{1}}\int_0^\infty\Psi_{\nu l0}^*(R,0,\varphi)\Psi_{\nu l'0}(R,0,\varphi)\dd{R}\\
    &= \frac{a\sqrt{(2l+1)(2l'+1)}}{4\pi V_{1}}{R}_{\nu l,\nu l'}^{(2)}\,.
\end{align}
Curiously, this radial matrix element vanishes when $l\ne l'$ \cite{Aguilar}: 
\be
{R}_{nl,nl'}^{(2)}= \frac{1}{\nu^3(l+1/2)}\delta_{ll'}.
\ee
This leads to a simple expression for the energy shifts,
\be
  \lim_{d\to 0}\tilde{V}_{lm,l'm'} =\frac{a}{V_{1}2\pi \nu^3}.
\ee
The Rydberg levels are identically affected by the scatterers and remain degenerate. Explicitly inserting the volume, we have 
\be
\tilde{E}^{1D}_{lm}=\frac{1}{ \pi\nu^5}.
\ee
The scaled energies scale as $\nu^{-5}$, decreasing slower with increasing $\nu$ than the $2$D-Composite energies. 

\subsubsection{3D scatterers}

For a homogeneous structure in $3$D the matrix elements are even simpler,
\begin{equation}
\begin{split}
    \lim_{d\to 0}\tilde{V}_{lm,l'm'} =\frac{a^3}{V_{3}}\int_V\Psi_{\nu lm}^*(R,\theta,\varphi)\Psi_{\nu l'm'}(R,\theta,\varphi)\text{d}^3R.
\end{split}
\end{equation}
This is the normalization integral, and thus all Rydberg states are again degenerate, but with a global shift, 
\be
\tilde{E}^{3D}_{lm} = \frac{3}{4\pi\nu^6}.
\ee
These scale as $\nu^{-6}$, more slowly than the $2$D-Composite. The Rydberg Composite spectrum thus obeys a power-law scaling behavior $\nu^f$, where $f=-5$ in 1D, $f=-11/2$ in 2D, and $f=-6$ in 3D. Despite the fact that the monolayer gives an energy scaling intermediate between the other geometries, it leads to a non-degenerate, highly structured dense limit which is totally distinct from the 1D and 3D-Composites.

\subsubsection{Interpolation of low and high band edge scaling for 2D scatterers}

We now explore the DoS scaling in the $2$D case in further detail to arrive at a universal DoS for 2D Rydberg Composites in the homogeneous limit.  We compute a smooth density of states,
\be
\frac{\delta N}{\delta \widetilde{E}} = \sum_{i=1}^N F(\widetilde{E};\sigma,\widetilde{E_i}),
\ee
where $F$ is a convolution function for the discrete data, i.e. a Gaussian or a box function centered at $\widetilde{E_i}$ and having width $\sigma$.  We focus first on the ``band'' region, where the eigenfunctions are to a good approximation labeled by integers $b$ and $k$, and which scale as $\nu^{-11/2}$. Since the number of states in the bands increases approximately linearly with $\nu$, we rescale the widths also so that they decrease linearly in $\nu$. Fig.~\ref{fig:DOS}a shows the resulting DoS for three $\nu$ values using a Gaussian distribution for $F$. The agreement between different $\nu$ is excellent in this band region, breaking down as energy increases. In grayscale we overlay each band separately, showing how the total DoS is built up from these, and in particular how the overlapping bands create the saturation point in the DoS and the eventual onset of the $\nu^{-6}$ scaling law. 
Fig.~\ref{fig:DOS}b shows a DoS with the $\nu^{-6}$ scaling, appropriate to the ``overlap region'' where the diagonal approximation breaks down. The widths now decrease as $\sigma/\sqrt{\nu}$ to obtain a smooth function. Some technical details involved in these figures are discussed in Appendix \ref{app:histograms}.

Fig.~\ref{fig:DOS}c presents a  scaling that smoothly interpolates between these two regimes as a function of $\widetilde{E}$. It allows us to construct  a ``universal'' density of states for the $2$D-Composite, independent of $\nu$. Details of this process, which uses a hyperbolic tangent to map the relevant scale factors, widths, and normalizations between these two regions as a function of $\widetilde{E}$, are provided in Appendix \ref{app:histograms}. The DoS shown in Fig.~\ref{fig:DOS}c confirm that this scaling is indeed universal, as the DoS for the three different $\nu$ levels are essentially indistinguishable.

\section{Evolution of the spectrum for decreasing density of scatterers}
\label{sec:evolution}

We have thus investigated the nature of the spectrum in the limit where the lattice cannot be resolved by the Rydberg wave function. For the $2$D-Composite this led to a non-trivial spectral density with band-like structures, scaling laws, and Rydberg wave functions quite different from the ones known from atoms or molecules. In this section we study  the characteristics of this system at lattice spacings large enough to be resolved by the Rydberg wave function. To this end we define a threshold lattice spacing, $d_\mathrm{c}$, below which the Rydberg wave function can no longer resolve the scatterers and which therefore replicates the homogeneous limit of scatterers, formally only reached for $d=0$.

\subsection{Transition to homogeneous density of scatterers}
Several circumstances complicate a rigorous definition of $d_\mathrm{c}$. First, the electron's wavelength varies spatially: in the radial direction it increases quadratically, while in the angular degree of freedom it is strongly $l$-dependent. Secondly, since the potential depends non-linearly on the wave function amplitude at the locations of the scatterers, it is not clear from the onset at what length scales scatterers can be resolved. 

 We have already seen that the homogeneous limit in the $2$D case is heralded by wave functions which are diagonal in $m$. Near the bottom of each band these states are also approximately diagonal in $l$. 
 %\blue{The $l$ value we select is $l=2 b-2+m$.}
 %\red{Using the band index and expression ($l=2 b-2+m$) from Sec. \ref{sec:monolayer} we can assign an $l$ to each state.}
 %Of this superposition, the maximal resolving power comes from the smallest $l$ state, i.e. when $m=l$.
 Such a wave function has $2m$ angular nodes in the plane, and hence has an angular resolution $\pi/m$. The quadratic scaling of the radial nodes implies that their density  decreases from the inner to the outer classical turning point. Therefore, the wave function can detect the smallest spatial features on a circle given by the inner classical turning point, $R_\mathrm{min}$. The (angular) resolution corresponds to the distance separating two adjacent nodes
 on this circle,  $w=R_\mathrm{min}\sin{(\pi/m)}$. Adapting $w$ to a square lattice gives a critical lattice spacing equal to $w/\sqrt{2}$. 
 %Setting $l=m$ to obtain the limiting case with maximum resolving power and dropping terms of order $\nu^{-2}$ we arrive at
 Dropping terms of order $\nu^{-2}$ and assuming the angular resolution is smaller than any radial wave function feature we arrive at a critical lattice spacing
\begin{equation}
    d_\mathrm{c}\approx\frac{\sin{(\pi/m)} \nu^2}{\sqrt{2}}\left[1 - \sqrt{1 - \frac{l(l+1)}{\nu^2}}\right].
    \label{andrew}
\end{equation}
%\begin{equation}
%    d_\mathrm{c}\sim\frac{\pi \nu^2}{\sqrt{2}(\nu-k-2b+2)}\left[1 - \sqrt{1 - \frac{(\nu-k)(\nu-k+1)}{\nu^2}}\right].
%    \label{andrew}
%\end{equation}

%
%To apply this to 
%Each energy level in the homogeneous limit of the $2$D-Composite can be labeled with $m$ and $\beta$, $E(\beta,m)$. 
%$l$ is needed to apply the $d_c(m,l)$ formula but is not well defined for each state. An approximate $l$ can be found using the expression $l=2 b-2+m$ where $b=\beta$ from Sec. \ref{sec:monolayer} to find $d_c(m,l)$ for each energy.
%We use the fact that each energy level $E_\beta(m)$ in the homogeneous limit is labeled by $m$ and a band index  $\beta$ (see Fig. \ref{fig:bandstates}), \red{ and the approximate expression $l=2 b-2+m$ from Sec. \ref{sec:monolayer}} to connect the band energies to these values $d_c(m,l)$. 
%Using this labeling the relationship $d_c(E)$ is obtained and plotted as  black curves in Fig.~\ref{fig:2d}. 
The black curves  in Fig.~\ref{fig:2d}  are lines through the points  $(d_{\beta,m}, E_{\beta,m})$,  where the connection between 
$\beta$ and $l$ is made with   the  approximate relation $l = 2(\beta-1) -m$, as discussed in Sec. \ref{sec:monolayer}.
These fit well with the qualitative transitions seen in the spectrum for the lower bands where the approximations are more accurate.
The transition can therefore be interpreted as the minimal spacing of scatterers which still can be resolved by the wave function.
This spacing $d_\mathrm{c}$ should not be confused with $d_{D}$ from Table I, which is the maximal spacing for breaking the degeneracy of all levels in the manifold $\nu$.
To keep the number of shifted states constant at $b_D = N_D$,
%Only for all levels participating
%in the interaction with the monolayer, the evolution with the fill factor $F$ or the lattice spacing $d$ is meaningful.
%Therefore, 
considerations for the remainder of this section will refer to $d\le d_{D}$.

\subsection{Between the homogeneous and the few-scatterer limit: Chaotic spectra}
\label{sec:chaos}

Random matrix theory (RMT) is an appropriate framework to analyze chaotic spectra.
Although the Rydberg Composite's spectra are in principle chaotic, the application of RMT to the present problem is hindered by the fact that for $d>d_c$ the system obeys several symmetry constraints when $F=1$.
Moreover, towards the ``trilobite-limit'' of only a few scatterers ($F\approx 0$ and or $d\gg d_\mathrm{c}$) the spectrum becomes regular. Both properties affect strongly the mean density of states. Hence, standard tools \cite{bohigas1990chaos} from RMT
to describe properties of a classically chaotic system  are cumbersome to implement as they require knowledge of the mean DoS for unfolding.
The unfolded DoS has uniform mean density \cite{AGR,RMT} 
and can be used to extract the
eigenvalue correlations in the spectrum.

\subsubsection{The adjacent gap ratio (AGR)}

To avoid unfolding, we resort to the so called 
%to analyze the energy spectrum for $F \ne 1$ and the inhomogeneous limit, random matrix theory (RMT) is used. RMT can be used if the classical counterpart of the system is chaotic \cite{AGR}. A robust measure used in RMT is the 
%
adjacent gap ratio (AGR) \cite{AGR_fermions,AGR}
 \begin{equation}
\text{AGR}=\left \langle \frac{\min(s_n,s_{n-1})}{\max(s_n,s_{n-1})}\right\rangle
\label{AGR}
\end{equation}
with $s_n=E_n-E_{n-1}$ and the average $\langle\rangle$ taken over the whole spectrum. When $F\ne 1$ we also average over many lattice realizations. 
%\red{needs reference to what unfolding means}
Since the AGR only depends on local fluctuations it does not require unfolding \cite{AGR}. AGR
can also deal with a ``mixed'' chaotic and regular spectrum, i.e., it  can  differentiate Poisson statistics, marking uncorrelated energies typically from preserved subspaces due to symmetries, from Gaussian Orthogonal Ensemble (GOE) statistics which occur for chaotic dynamics \cite{bohigas1990chaos} without additional symmetries, when level repulsion is present  \cite{RMT}. As RMT references  for regular and chaotic Rydberg Composite dynamics we obtain,
for a matrix size corresponding to the $\nu=30$ case, AGR$_\mathrm{P}=0.386$ and AGR$_\mathrm{GOE}=0.530$ for Poisson statistics and GOE statistics, respectively. For  further details see Appendix \ref{app:AGR}.

 \subsubsection{The evolution of AGR with the fill factor for different fixed lattice spacings}
 
One can see in Fig.~\ref{fig:NAGR}a  that towards small fill factors, but compatible with $N>N_{D}$  where the spectrum looks chaotic, the AGR function  $g_{d}(F)$ indeed approaches  $g^{0}\equiv$~AGR$_\mathrm{GOE}$ for all $d$ shown, see Appendix \ref{app:AGR}. However,  $g_{d}(F)$ breaks off $g^{0}$ for increasing $F$, to reach eventually the value $g_{d}(1)=0$  due to  geometry induced degeneracy. 
%The break-off occurs at larger $F$ as  larger $d$ is.
For larger $d$ the break-off occurs at larger $F$.
For large $d$ the AGR function is  box-like in shape with a sudden transition to $g_{d}(1)=0$. We note here in passing that to a good approximation the family of AGR functions $g_{d}(F)$ shown follow the form
\begin{equation}\label{agr-scaling}
(g_{d}/g^{0})^\gamma+F^\gamma=1,\,\,\,\,\,\, \gamma = 1+d/2\,,
\end{equation}
an interesting relation revealing a self-similar property, whose deeper analysis is beyond the scope of this work. 
%  We recall that every energy is doubly degenerate in that case, and hence AGR is trivially $0$, as can be seen from Eq. \eqref{AGR}. On the other hand, as long as  AGR starts to decrease gently at  lower $F$ as the \JM{This colored bit will change shortly as a universal curve is being made! filling factor represents a total number of scatterers which translates within the area of the wavefunction intersecting the atomic layer into filled lattice with effective spacing $d_\mathrm{eff}$  eventually smaller than $d_\mathrm{c}$  ($d_\mathrm{eff}=d_\mathrm{c}$ is shown with arrows 
%  for the respective lattice constants $d$). }
%  We can accommodate this degeneracy in the $d\to0$ case by  considering only the $m\ge 0$ states to evaluate the AGR. Then we  obtain AGR$_{m\ge0}=0.5279$ in good agreement with the GOE value, 
%revealing that this geometric degeneracy in $m$ is the only one left in the homogeneous limit. 

  \subsubsection{The evolution of AGR for a filled lattice with decreasing lattice spacing}

\begin{figure}[t]
     \centering
     \includegraphics[width=\columnwidth]{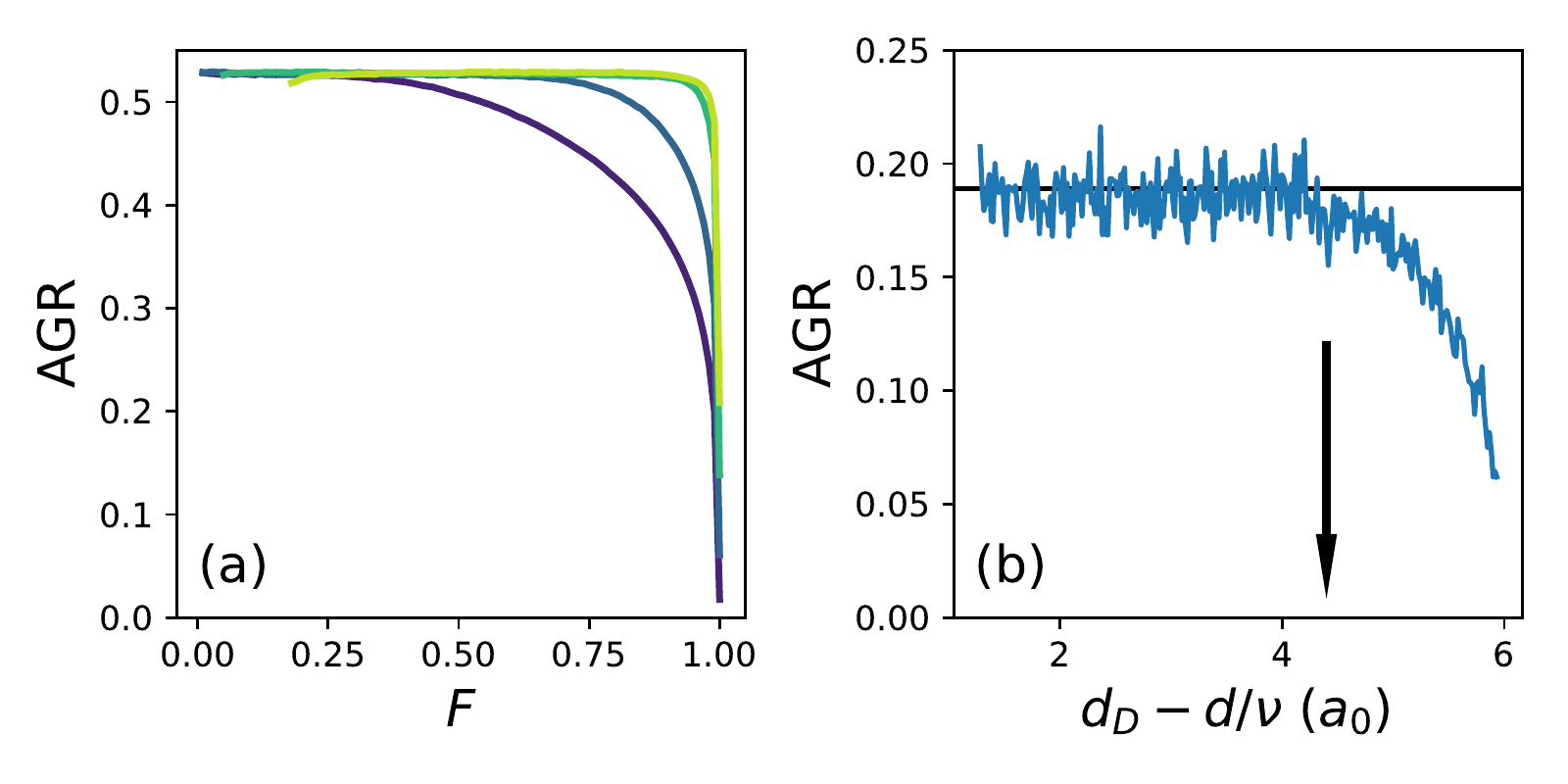}
     \caption{(a) Average AGR of 2000 realizations of a lattice of scatterers as a function of $F$ for $d=5,10,30$,and $70$ in sequence from dark to lighter line color, $\nu=30$.
     (b)  AGR for a filled lattice as a function of $d$. The arrow marks $d_\mathrm{c}$ while the horizontal lines show the AGR value found using different configuration of random matrices to model the homogeneous and high $d$ limit, see text.}
     \label{fig:NAGR}
 \end{figure}

For a filled lattice $F = 1$,  Fig.~\ref{fig:NAGR}b reveals that neither GOE nor Poisson values match with the statistics observed  for any lattice spacing $d$. For small $d$ towards the homogeneous limit, the AGR approaches zero again due to the geometrically induced degeneracies. However, even for $d>d_\mathrm{c}$ the AGR settles to a value different from the GOE one due
to the inherent symmetries of our system. We have simulated a chaotic system obeying the inherent ``crystal'' symmetries  by a block diagonal GOE matrix with each block representing one irreducible representation of the $C_{4v}$ point group (see Appendix \ref{app:AGR}). This synthetically obtained AGR (black line in Fig.~\ref{fig:NAGR}b)
agrees well with the AGR of the true spectrum for $d>d_\mathrm{c}$.

We may conclude that the spectral fluctuations in the DoS are indicative of a chaotic system with symmetries separating states into none interacting blocks in the $F=1$ limit, while for $F\ne 1$ these symmetries gradually break until the spectrum is purely chaotic.

\section{Experimental realization}
\label{sec:experiment}
\subsection{Isolation of Rydberg manifolds in the presence of scatterers}

Rydberg Composites live in the Hilbert space of a single Rydberg manifold $\nu$. This implies that the excitation must be high enough such that interactions with adjacent manifolds is negligible. We have shown that
the normalized energy levels of  Rydberg Composites in the dense lattice limit scale as $\nu^{-5}$, $\nu^{-11/2}$, and $\nu^{-6}$ in $1$D, $2$D, and $3$D, respectively. These  energies must be compared, as a function of $\nu$, with the overall spacing between Rydberg manifolds in order to ascertain the isolation of the Rydberg Composite's manifold.
 %if it is a reasonable approximation to truncate to only a single Rydberg manifold.
  The Hellman-Feynman theorem guarantees that the coupling between energy levels increases inversely to their energetic separation, and so we must confirm that the Composite spectrum does not overlap, or even approach, an adjacent Rydberg manifold. The spacing between Rydberg manifolds decreases as $\nu^{-3}$, and hence the scaling of the un-normalized Rydberg Composite spectra must fall faster than this value. In $1$D, the un-normalized spectrum is:
\be
E_{1}\sim 2\pi |a_s|\tilde{V}_{1}\frac{\nu^2}{d}\frac{1}{2\pi\nu^{5}} \propto \frac{ |a_s|}{\nu^4},
\ee
since $d\propto\nu$. In $2$D, we have (for the strongest scaling,  $\nu^{-11/2}$):
\be
E_{2}\sim2\pi |a_s|\tilde{V}_{2}\frac{\nu^4}{d^2}\frac{1}{\pi^{3/2}\nu^{11/2}}\propto\frac{2|a_s|}{\sqrt{\pi}\nu^{7/2}},
\ee
again, using $d\propto\nu$. Finally, in $3$D, $d\propto\nu^{4/3}$, and so
\be
E_{3}\sim2\pi|a_s|\tilde{V}_{3}\frac{\nu^6}{d^4}\frac{3}{4\pi\nu^6}\propto\frac{3}{2}\frac{|a_s|}{\nu^4}.
\ee
In all three cases the Rydberg Composite's energies decrease faster than the splitting between manifolds as a function of $\nu$, and hence at sufficiently high $\nu$ only the states of a single manifold contribute and the Rydberg  Composite exists as described.

\subsection{Experimental choice of $\nu$}
The requirement for isolation of the Rydberg Composite's manifold $\nu$ stretches certain experimental possibilities and therefore, it might be desirable to choose a scatterer species with a smaller scattering length than Rb or Cs, the current standards. For example, sodium ($a_s(0)\sim -5$) and lithium ($a_s(0)\sim -7$) have scattering lengths only a third to a half that of Rb \cite{EilesHetero}.

\begin{figure}
     \centering
     \includegraphics[width=\columnwidth]{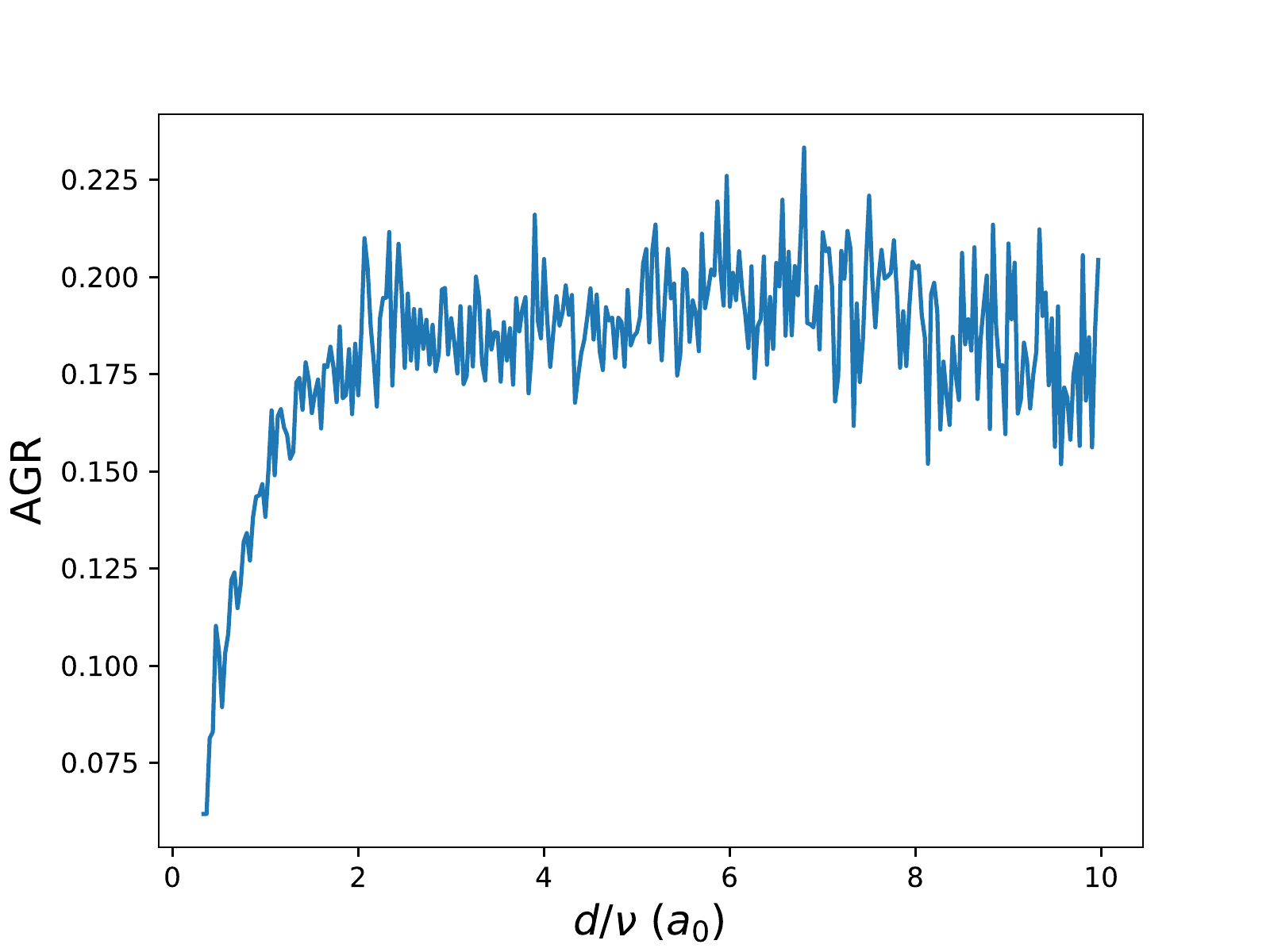}
     \caption{AGR for a filled lattice for a full lattice as a function of $d$.}
     \label{fig:AGR_big}
 \end{figure}

A high $\nu$ is also important in order to reach an experimental regime where our additional approximations -- constant scattering length and vanishing inter-atomic potentials -- are realistic. The model Hamiltonian of Eq.~\ref{eq:ham} neglects scattering contributions from higher partial waves and  the polarization interaction between the Rydberg core and the scatterers.
%and any effect of the potentials used to structure the environment (such as an optical lattice). 
All of those approximations become more accurate at higher $\nu$.
 The most serious obstacle is probably the current experimental capability in creating small lattice spacings. If the experiment was performed in an optical lattice this would require quite a large principal quantum number, as the current minimum lattice spacing is around $d= \lambda /6 \sim 2500a_0$ \cite{small_lattice}. However, it may be possible to observe experimentally the onset 
 of the chaotic behavior of the Rydberg Composite spectrum as discussed in section \ref{sec:chaos} when shrinking the lattice spacing as far as possible.
 This chaotic behaviour continues to larger lattices spacings than $d_D$ 
 %for the lower parts of the spectra that still experience a shift 
 . 
 %The AGR value stays near the high $d$ chaotic value up to $d\sim 10\nu$, this in combination with the minimum optical lattice spacing, would require $\nu\sim 250$ to observe. 
 One can see in Fig. \ref{fig:AGR_big} that at the ratio $d/\nu=10$ the chaotic AGR value is still
 present; at the minimum optical lattice spacing mentioned above this
 %present, this, in combination with the minimum optical lattice spacing, 
 would require $\nu\sim 250$ to observe.
 This is feasible given that
 Rydberg states with $\nu\sim 300-500$ have been produced \cite{huge_n}. 
% putting measurement of this regime on the edge of experimental feasibility.
 
 %\red{need discussion here about optical lattice spacings, wavelengths, tweezer possibilities, solid state aspects...?}

\section{Conclusions and future work}
\label{sec:conclusion}
In this article we have introduced  Rydberg Composites built by coupling a Rydberg atom to a dense distribution of many neutral atoms immersed within the Rydberg wave function. Rydberg Composites provide a systematic interpolation from the trilobite and polyatomic few-body regime to a dense environment with a homogeneous density of scatterers as the asymptotic limit. 
Rydberg Composites, particularly the $2$D monolayer case emphasized here, are a new form of matter intersecting few-body atomic Rydberg physics, quantum dynamics involving optical lattices, and few-body quasiparticle examples from solid state physics. 

One can imagine many immediate possibilities to extend this concept. These include more refined geometries of the
embedding environment to tune the Rydberg Composite spectrum, a goal which is traditionally reached by applying
external electric or magnetic fields.
%The scatterer distribution could be modified extensively: for example multiple $1$D lattices, stacked monolayers of identical or different atomic species, hybrid-dimensional lattices such as a rectangular lattice, or distributions more closely connected to modern BEC experiments such as narrow cigar distributions or distributions with non-uniform density. The effect of the lattice nature on the resulting eigenfunctions and lattices can be explored as well, for example to determine how the properties of a honeycomb lattice differ from a square lattice such as we studied here, and following that how the properties are expected to change if the lattice is replaced with a material such as graphene.
Also, localization and decoherence studies are feasible by removing the frozen gas restriction and either shaking the lattice explicitly or allowing it to move randomly at some finite temperature. 

\appendix

\section{Matrix representation in the ``trilobite basis''}
\label{app:trilobitemethod}
In the context of polyatomic Rydberg molecules it has previously proven  useful to perform a change of basis from the manifold of Rydberg states $|\nu lm\rangle$ to the basis of trilobite \textit{dimer} states $|i\rangle$ \cite{eiles2019trilobites}. Each element of this non-orthogonal basis is a trilobite wave function extending from the Rydberg core to the $i$th scatterer, 
\be
\label{eq:trilobasisdef}
\Upsilon(\vec R_i,\vec r) = \sum_{lm}\phi_{\nu lm}^*(\vec R_i)\phi_{\nu lm}(\vec r). 
\ee
This basis is ideal when $\nu^2\gg M\gg 1$, as it greatly reduces the numerical challenges associated with the large $\nu^2$-dimensioned Rydberg basis. It furthermore provides significant qualitative insight into the structure and possible geometries of polymers since the eigenstates give directly the contribution of each scatterer within the configuration to that energy configuration. From this one can define alternative localization measures utilizing the information immediately available from these eigenvectors, or classify the states within this basis using the known symmetries of the scatterer configurations, as we do in Appendix \ref{app:symorbitals}. 

In this appendix we extend this method to the Rydberg Composite. Since this system typically has $M\gg \nu^2$, the trilobite basis is no longer numerically beneficial. It can still provide useful qualitative insight, and in the infinite scatterer limit it leads to an alternative method, solving an integral equation, to obtain the spectrum. 

Within our stated approximations, the representation of $H$ in the trilobite basis is the $M\times M$ matrix $H_{ii'}=\Upsilon(\vec R_i,\vec R_{i'})$. One numerical advantage of this approach is that the matrix element $H_{ii'}$ can be expressed using only $u_{n0}(R)$ and $u_{n0}'(R)$,  eliminating the need to evaluate many high-$l$ wave functions when $\nu\gg 1$ \cite{EilesJPB}. If $M>B_D$ the diagonalization of $H_{ii'}$ in this representation yields $M- B_D$ vanishing eigenvalues in addition to the $B_D$ shifted eigenvalues, and one numerical disadvantage lies in distinguishing these from real, but small, eigenvalues.  As $M\to\infty$, the dimension of $H_{ii'}$ becomes infinite, and hence the eigenvalue equation becomes an integral equation,
\be
\Psi(\vec r) = \frac{1}{\tilde{E} }\int_\mathcal{V}\Upsilon(\vec R,\vec r)\Psi(\vec R)\dd{\mathcal{V}},
\label{eq:inteq}
\ee
where $\mathcal{V}$ is the scatterer volume in dimension $D$. Since $\Upsilon(\vec R,\vec r)$ is separable, Eq. \ref{eq:inteq} has solutions when $\tilde E$ is obtained from the determinantal equation
\be
0=\det \left|\delta_{ii'} - \frac{1}{\tilde{E} }\int_\mathcal{V}\Psi_i^*(\vec R)\Psi_j(\vec R)\dd{\mathcal{V}}\right|.
\ee
This equation can be solved via a numerical root finder.

\section{Parabolic coordinates}
\label{app:parabolic}
The hydrogen atom separates in many coordinate systems, and one should choose a coordinate system that is, if possible, adapted to the geometry of the scatterer distribution. For example, the eigenstate for a single scatterer is nearly proportional to a Rydberg basis wave function in ellipsoidal coordinates \cite{Granger}, and the scatterer operator in the dense lattice limit for $1$D and $3$D-Composites commutes with the Hamiltonian in spherical coordinates. In the $2$D case, the spherical wave functions are clearly not well adapted to the scatterer distribution.  Although cylindrical coordinates are well-suited to the $2$D-Composite scatterer distribution, the Coulomb potential does not separate in these coordinates. It does, however, separate in parabolic coordinates,  
\be
x = \sqrt{\xi\eta}\cos\phi,y=\sqrt{\xi\eta}\sin\phi,z = \frac{1}{2}(\xi-\eta).
\ee
These treat parabolas $\xi$ and $\eta$ on either side of the $z=0$ plane democratically, and therefore could be more closely adapted to the $2$D-Composite. As the following shows, the Hamiltonian in this coordinate system still must be numerically diagonalized, although it does have closed-form analytic matrix elements using the analytic forms for the hydrogen wave functions in parabolic coordinates. We therefore present this calculation not for its direct usefulness to the problem at hand, but to define a potentially useful yet uncommonly employed starting point that could benefit future calculations of the properties of Rydberg Composites. 

The hydrogen wave function in these coordinates is
\begin{align}&
\Psi_{n_1,n_2,m}(\xi,\eta,\phi)=\frac{e^{im\phi}}{\sqrt{n\pi}}e^{-\frac{1}{2}\beta(\eta + \xi)}(\eta\xi)^{|m|/2}\\&\times(n_1+|m|)!(n_2+|m|)!L_{n_1}^{|m|}(\beta\xi)L_{n_2}^{|m|}(\beta\eta)\nonumber
\\&\times\nonumber\sqrt{\frac{n_1!n_2!}{((n_1+|m|)!(n_2+|m|)!)^3}}\beta^{|m|+3},
\end{align}
where $n = n_1 + n_2 + |m| + 1$ and $\beta = \frac{1}{n}$.
This is normalized with respect to the volume element, $(\xi + \eta)/4\dd\xi\dd\eta\dd\phi$. The matrix elements of the scatterer potential are
\begin{align}
&V_{n_1n_2m,n_1'n_2'm'}=2\int_{0}^\infty\int_0^\infty\int_0^{2\pi}\delta(\eta-\xi)\\&\Psi_{n_1n_2m}(\xi,\eta,\varphi)\Psi_{n_1'n_2'm'}(\xi,\eta,\varphi)
\frac{\xi+\eta}{4}\dd{\eta}\dd{\xi}\dd{\varphi},\nonumber
\end{align}
using
\be
\delta(z)=\delta\left(\frac{1}{2}(\xi-\eta)\right)=2\delta(\xi-\eta).
\ee
Integration over $\varphi$ again leads requires $m=m'$, while integration of $\xi$ sets $\xi=\eta$. The resulting expression involves only an integral over $\eta$,
\begin{align}
&\langle n_1n_2m|V|n_1'n_2'm'\rangle = 2\pi\delta_{mm'}\\&\times\int\eta\Psi_{n_1,n_2,m}(\eta,\eta,0)\Psi_{n_1'n_2'm'}(\eta,\eta,0)\dd{\eta}.\nonumber
\end{align}
The allowed quantum numbers are also restricted such that $n_1 + n_2 = n_1' + n_2'$ since $m$ is conserved and $n$ is the same within a single manifold expansion. The general integral of this form has a closed form solution,
\begin{widetext}
\begin{equation}
\begin{split}
&\int_0^\infty e^{-2x/n}x^{2m+1}L_{n_1}^{m}(x/n)L_{n_2}^{m}(x/n)L_{n_1'}^{m}(x/n)L_{n_2'}^{m}(x/n)\dd{x}=  \frac{1}{(-1)^{4m}(n_1+m)!(n_2+m)!(n_1'+m)!(n_2'+m)!}\\&\times\sum_{i=m}^{n_1}\sum_{j=m}^{n_2}\sum_{k=m}^{n_1'}\sum_{l=m}^{n_2'} b_{in_1+m,m}b_{jn_2+m,m}b_{kn_1'+m,m}b_{ln_2'+m,m}\frac{(i+j+k+l+\beta-4m)!\lambda^{i + j + k + l - 4m}}{\alpha^{i+j+k+l+\beta - 4m+1}},
\end{split}
\end{equation}
\end{widetext}
where
\be
b_{inm}^{\lambda}=\frac{\lambda^{i-m}(-1)^i(n!)^2}{(n-i)!(i-m)!i!},
\ee
and
$\beta = 2m+1$, $\alpha = 2/(n_1 + n_2 + m + 1)$, and $\lambda = \alpha/2$. Thus we obtain $(2m+1)\times(2m+1)$ block diagonal matrices since $n_1$, $n_2$, are related to $n$ and $m$. Diagonalization of these matrices then yields the spectrum computed in the text. Since the basis was not restricted to reject wave functions with no amplitude in the plane from the beginning, as we did in spherical coordinates, $\nu(\nu-1)/2$ of these eigenvalues vanish.
\section{Symmetry adapted orbitals}
\label{app:symorbitals}
In this appendix we briefly review the use of the projection operator method which, in conjunction with the trilobite basis representation developed in Appendix \ref{app:trilobitemethod}, can be used to obtain the Rydberg Composite spectrum when the scatterer configuration is a member of a molecular point group. The particular utility of this approach is that it leads to a classification of the resulting degeneracies and level crossings in the spectrum in the finite lattice-size regime. The description here follows Ref. \cite{EilesJPB} and is valid only for the $s$-wave (contact potential) interactions used here; generalization to $p$-wave interactions requires additional complications \cite{EilesJPB}. We obtained the symmetry-adapted eigenstates via the following process:
\begin{itemize}
    \item Identify the relevant molecular point group. For example, the planar square lattice satisfies the $C_{4v}$ molecular point group. 
    \item Construct the labelled basis of trilobite-functions, $\vec v$, where $v_k = \Upsilon(\vec R_k,\vec r)$. 
    \item Every symmetry operator in the point group corresponds to a rotation/reflection matrix, denoted $\underline{r_i}$. This operator acts on the position vector $\vec R_k$ of each trilobite function in $\vec v$, changing it to a different position vector, i.e. $\underline{r_i}\vec R_k  = \vec R_j$. 
    \item With this information, define an operator $\mathcal{R}_i$ which acts not on the position vectors $\vec R_k$ but rather on the basis vector $\vec v$. Its elements are $\left(\mathcal{R}_i\right)_{jj'}=\delta_{jk}\delta_{j'k'}$, where $k$ and $k'$ are related by $\underline{r_i}\vec R_{k'} = \vec R_k$. 
    \item Eq. 20 of Ref. \cite{EilesJPB}, in conjunction with the point group's character table, yields the projection operators $\hat{\mathcal{P}}^j$. These are (when properly rank-reduced) $M_j\times M$-dimensioned matrices, where $M_j=\text{Tr}(\hat{\mathcal{P}}^j)$. These traces satisfy $\sum_jM_j = M$, and thus describe how the total number of eigenstates are partitioned into each irreducible representation. 
    \item These projection operators are then used to partition the Hamiltonian $H_{ii'}$ into block-diagonal form, where each block $H_{kk'}^j$ is the reduced Hamiltonian for the $j$th irreducible representation. This is done via the transfomration
    \be
    H_{kk'}^j = \sum_{i=1}^M\sum_{i'=1}^M \mathcal{P}^j_{ki}H_{ii'}\left(\mathcal{P}^j\right)^\dagger_{i'k'}
    \ee
\end{itemize}
Finally, each $H_{kk'}^j$ is diagonalized. The eigenstates of a given $j$ exhibit avoided crossings when a parameter, such as $d$, changes, while the eigenstates corresponding to different irreducible representations (different $j$ values) exhibit real crossings. To make this concrete, we see for the $C_4$ symmetry of the plane that exactly half (neglecting ``round-off'' errors due to the mismatch between lattice points in the square and the circular Rydberg orbit) of the eigenstates are in the $2$D $E$ irreducible representation, while the remaining 50\% of the eigenstates are approximately evenly split among the remaining irreducible representations, $A_1$, $A_2$, $B_1$, and $B_2$.
\section{Further details on the smooth DoS}
\label{app:histograms}
This appendix describes additional technical details regarding the density of states calculated in Sec. \ref{sec:homogeneous}. We begin with the full expression for the density of states used to make Fig. \ref{fig:DOS}a and b,
\be
\label{DOS1}
\frac{\delta N}{\delta \widetilde{E}} = \frac{1}{\nu^{1/g}b^2}\sum_{i=1}^{B_D}F(\widetilde{E};\frac{\sigma b}{\nu^g},\widetilde{E_i} b \nu^{f}).
\ee
In this formula, $F(x;\sigma,x_i)$ is a function to convolve the discrete line spectra with a finite width distribution.
%represent the discrete line spectra with states of finite width.
In Fig. \ref{fig:DOS} a Gaussian function 
\be
F(x;\sigma,x_i) = \frac{1}{\sqrt{2\pi\sigma^2}}e^{-\frac{(x-x_i)^2}{2\sigma^2}}
\ee
was used. This has unit normalization, width $\sigma$, and peaks about its mean, $x_i$. Other functions, e.g. box functions, could be chosen as well. As discussed in the text, there are different scaling laws for the width $\sigma$ and energy levels $x_i$ for the different regions - ``band'' and ``overlap'' - of the density of states. Specifically, in the ``band'' region this is handled by setting $f = 11/2$, $g = 1$, and $b = 1$. We found that $\sigma=0.1$ sufficed to achieve the smooth resolution of Fig. \ref{fig:DOS}a. The integrated density of states is
\be
N_\text{band} = \frac{B_D}{\nu}
\ee
In the ``overlap'' region we set $f = 6$ and $g = 1/2$; this rescaling of the widths is necessary since the overlap states are denser. The integrated DOS in this case is
\be
N_\text{overlap} = \frac{B_D}{\nu^2\cdot b^2}.
\ee
One inelegant technical detail stems from the fact that the band and overlap regions span very different energy ranges due to the difference between the $\nu^6$ and $\nu^{11/2}$ scale factors. As a result, we must apply a global compression of the ``overlap'' energies by multiplying by a somewhat arbitrary factor, $b$, and afterwards normalize the amplitude of the overall expression with a factor $b^{-2}$. We find that $b=0.1$ sets, for this range of $\nu$, the two scaled DOS to lie between the same ordinate and abscissa limits.

The fully universal scaling of the whole DOS is accomplished by making $b$, $f$, and $g$ functions of $\widetilde{E}$:
\be
\frac{\delta N}{\delta \widetilde{E}} = \frac{1}{\nu ^{1/g(\widetilde{E})}b(\widetilde{E})^2}\sum_{i=1}^{N} F(\widetilde{E};\frac{\sigma b(\widetilde{E})}{\nu ^{g(\widetilde{E})}},\widetilde{E_i} b(\widetilde{E}) \nu ^{f(\widetilde{E})}). 
\ee
For each of these fit functions we have found that a tanh function is sufficient to interpolate between band and overlap regions.  spanning the range from $v_1$ to $v_2$ with a width $w$ and center $x_0$ provides a smooth interpolating function to transition between these two regions once these parameters are fit to the data. 
\be
f(x) = v_2 + \frac{v_1}{2}\left(1 + \tanh\left(\frac{x-x_0}{w}\right)\right)
\ee
For the case shown in Figs. \ref{fig:DOS}, $x_0 = -0.011$ and $w = 0.0028$. 
\section{AGR}
\label{app:AGR}
The random matrix AGR values were all calculated by diagonalizing 2000 realizations of real symmetric matrices whose matrix elements were randomly sampled from a normal distribution. 
%The GOE and Poisson values have been previously calculated \cite{AGR_fermions,AGR} with different values being found for the GOE case.
We observed the AGR values have a weak dependence on matrix size and hence calculated the values used in this paper on matrices of size 465, corresponding to the $\nu=30$ case for which the majority of our numerical data was calculated.

%\subsection{GOE and Poisson}

The AGR value for the GOE case was found to be AGR$_{GOE}=0.5304 \pm 0.0003$ 
using dense random matrices. The Poisson value was found to be AGR$_\mathrm{P}=0.3864\pm 0.0003$ using a random matrix with elements only down the diagonal.

%\subsection{GOE with symetries}

In the case of $F\approx1$ the symmetries of the system need to be taken into account. There are two limiting cases, the homogeneous and large $d$ case. The AGR value in the homogeneous case is trivially zero since each case is doubly degenerate due to the $\pm m$ symmetry. The system in the large $d$ case belongs to the $C_{4v}$ symmetry group. The $C_{4v}$ character table has 5 different types of irreducible representation in it and only states in the same irreducible representation can interact with one another. One of the irreducible representations is two-dimensional, meaning that each state belonging to it is doubly degenerate.

%For the large $d$ limit 
The AGR value for the large $d$ case is calculated using matrices constructed of six GOE matrices in a block diagonal format. 
The size of each block is chosen to match the number of states in each symmetry irreducible representation found in Appendix \ref{app:symorbitals}.
Four of the blocks are of size 56 (to account for the one dimensional irreducible representations) while the last two are identical (to account for the two-dimensional irreducible representation) and of size 120 each. From this we obtain a value of $AGR_{d_{large }}=0.1894 \pm 0.0002$. 

\nocite{apsrev41Control}
\bibliographystyle{apsrev4-1}
\bibliography{bib.bib}

\end{document}